# Ion-Containing Bottlebrush Elastomers as Pressure-Sensitive Electroadhesives

Hao Dong[1#], Intanon Lapkriengkri[1,2#], Nadia Chapple[2], Hyunki Yeo[1,2], Alexandra Zele[2], Hiba Wakidi[1,3], Thuc-Quyen Nguyen[1,3], Michael L. Chabinyc[1,2], Christopher M. Bates[1,2,3,*], Megan T. Valentine[1,4,*]

[1]Materials Research Laboratory, University of California Santa Barbara, Santa Barbara, CA 93106, USA. [2]Materials Department, University of California Santa Barbara, Santa Barbara, CA 93106, USA. [3]Department of Chemistry and Biochemistry, University of California Santa Barbara, Santa Barbara, CA 93106, USA. [4]Department of Mechanical Engineering, University of California Santa Barbara, Santa Barbara, CA 93106, USA.

\# These authors contributed equally to this work.
\* E-mail: cbates@ucsb.edu, valentine@engineering.ucsb.edu

**Key words:** bottlebrush polymer, pressure-sensitive electroadhesive, switchable adhesion

## Abstract

This study presents a materials-design framework for low-voltage pressure-sensitive electroadhesives based on ion-containing bottlebrush polymers that combine the on-demand reversibility of traditional electroadhesives with the tunable conformability typical of pressure-sensitive adhesives (PSAs). Two complementary bottlebrush polymers bearing pendant flexible side chains and independently tunable anionic or cationic groups were designed to form soft and tough elastomers after crosslinking. When the two oppositely charged bottlebrush networks were brought into contact, a smooth, continuous interface formed, which is locally charge neutral due to the presence of mobile counterions. At low voltages ($\leqslant$ 2 V), mobile ions migrate toward the electrodes, creating an interfacial heterojunction and significant electrostatic attraction that enhances adhesion, yielding an on/off ratio of up to more than 4.5. The low-voltage operation and PSA-like mechanics of bottlebrush electroadhesives, even at charge density as low as 18 C/g, create opportunities in applications such as soft robots, haptic devices, and biomedical devices.

# Introduction

The ability to control when and where adhesive bonds form remains critical in the design of advanced materials and devices, including soft robots[1, 2], biomedical implants[3, 4], and haptic interfaces[5]. Beyond contributing to performance metrics during use, the concept of switchable adhesion is also attractive in the context of sustainability and end-of-life considerations such as recycling and material recovery[6, 7]. Switchable adhesion has been demonstrated using a variety of stimuli including heat[8, 9], light[10], pH[11], and electricity[12-14]. Among these various triggers, electricity has a number of distinct advantages. Unlike temperature effects, which are often sluggish due to polymer dynamics[8], and light[8, 15] or pH-sensitive functionality[16] that requires significant penetration/diffusion to be effective, electrical stimulation can enable relatively rapid and precise control when electrodes can be readily incorporated into the system. The advantages of so-called "electroadhesives" have already been demonstrated in a wide range of emerging applications, from the treatment of inflammation[3] to robotic clutches[17] and exoskeletons[18].

A common class of electroadhesives relies on dielectric effects[14, 19] in an insulating layer overlying a patterned electrode[1, 14, 20, 21]. When voltage is applied across the electrodes, the dielectric layer polarizes and induces bound surface charge on the electrodes, producing an attractive electrostatic interaction and thus an adhesive force. The voltage required to achieve an electric field resulting in a significant change in adhesion depends on the thickness of the dielectric layer; this factor leads to challenges in lowering the voltage without compromising the ability of the adhesive to deform over the topography of a given surface. In practice, to maintain a reasonable thickness of the electroadhesive layer, the required voltage often exceeds 1 kV. Such extreme electric field strengths[22] (tens of MV/m) are close to the dielectric breakdown of polymers, leading to irreversible device failure and raising significant safety concerns[23]. Although in principle these materials provide value in devices like grippers and crawling soft robots, the necessary voltages are impractical in most real-world applications.

Ionic polymers have recently emerged as an alternative strategy to create low-voltage electroadhesives. For example, ionic hydrogels adhere to a range of materials under a DC voltage of approximately 10 V, although the precise mechanism of adhesion remains unclear[4, 24, 25]. Recently, a complementary design dubbed polymer heterojunctions[13, 26, 27] has demonstrated promising performance when two polymeric ionic liquids (PILs) possessing opposite charges, along with both positive and negative mobile counterions, are brought into adhesive contact[13]. When a sufficient voltage is applied, mobile ions migrate toward each electrode and expose charges attached to each polymer backbone at the polymer–polymer interface, which generates an attractive force across the junction[13]. Removing the voltage causes mobile ions to again redistribute within the polymer matrices and the electrostatic force disappears.

The significant decrease in voltage required to operate a well-designed ion-containing polymeric electroadhesive suggests this class of materials provides potential advantages in switchable adhesive applications. The polymer heterojunction design is particularly appealing because the all-solids formulation contains no solvent such as water or plasticizer that can leach out and change adhesive performance over time. However, the need to incorporate a fairly high density of ions introduces a unique constraint. Ions generally raise the glass-transition temperature (and/or crystallinity) of a polymer, making these electroadhesives rather stiff and more reminiscent of structural adhesives. There are, however, other equally important classes of adhesives that represent a significant share of the global market. One example is soft and conformable pressure-sensitive adhesives (PSAs), such as those found in stickers, tapes, and labels. How might one design a pressure-sensitive electroadhesive if the ions critical for performance inherently change the mechanical properties in ways that undermine pressure-sensitivity?

Here, we address this challenge by designing pressure-sensitive electroadhesives based on bottlebrush elastomers. The bottlebrush architecture—featuring a long, polymeric backbone with polymeric side chains protruding from each repeat unit—is well-known to reduce the modulus of a crosslinked elastomer by minimizing entanglements. We demonstrate that bottlebrush elastomers synthesized with rubbery poly(4-methylcaprolactone) (P4MCL) or poly(dimethylsiloxane) (PDMS) side chains and oppositely charged ionic comonomers form pressure-sensitive electroadhesives that are soft, conformable, and effective, even with ion densities one order of magnitude lower than the heterojunctions described above, despite having a similar number of ionic groups per polymer. Adhesion and deadhesion occur within seconds under an applied electric field, and optimal molecular designs yield an on/off ratio (defined as the ratio of adhesion with the voltage on versus off) greater than 4.5—exceeding linear poly(ionic liquids) at voltages as low as 2 V. In summary, bottlebrush-based electroadhesives represent a promising platform for expanding the scope and impact of electrically responsive and switchable adhesives.

## Results and Discussion

### Design of ion-containing bottlebrush elastomers

An ideal pressure-sensitive electroadhesive should rapidly respond to an external voltage while maintaining the mechanical properties of common PSAs. The unique bottlebrush architecture facilitates this design by effectively reducing the entanglement and counteracting the stiffness imparted by covalently bound ions. In particular, flexible side chains with a low glass-transition temperature ($T_g \ll 25$ °C) promote conformal bonding that is characteristic of pressure-sensitive adhesives, even after attaching ionic groups to the backbone. Importantly, the number of ionic groups can be tuned independently, while mobile counterions locally neutralize these tethered charges and simultaneously enable responsiveness to externally applied electric fields. This contrasts current linear poly(ionic liquid) systems[13, 26, 27], for which the charge fraction (that is, the mole fraction of ionic monomers) is coupled to the fraction of crosslinkers, so varying the charge content may also change the crosslink density. This coupling limits independent control over the charge density of the polymer network. Accordingly, we synthesized a complementary pair of bottlebrush polymers—poly[norbornene-polydimethylsiloxane]-*stat*-poly[norbornene-carboxylate]$^-$K$^+$ (BB-Anion, Fig. 1, left) and poly[norbornene-poly(4-methylcaprolactone)]-*stat*-poly[norbornene-imidazolium]$^+$I$^-$ (BB-Cation, Fig. 1, right)—via ring-opening metathesis polymerization (ROMP, see details in Supporting Note 1, Fig. S1–14 and Table S1). As shown in Fig. 1a, the backbone degree of polymerization ($N_{BB}$) of both BB-Cation and BB-Anion was set to be 100 (theorical value, see Supplementary Note 1 for details). The macromonomer side-chain degree of polymerization ($N_{SC}$) was 10 for BB-Cation and 74 for BB-Anion as measured by $^1$H nuclear magnetic resonance and gel-permeation chromatography. The architectures of BB-Cation and BB-Anion also provide a means to tune the charge fraction. The charge fraction $\alpha$ of each sample is defined as $\alpha = y / (x + y)$, where $x$ is the number of flexible side chains and $y$ is the number of ionic repeat units per bottlebrush molecule (see Supplementary Note 1 for details). For BB-Cation, the nominal charge fraction $\alpha$ ranged from 0% to 24%, whereas for BB-Anion $\alpha$ ranged from 43% to 51%.

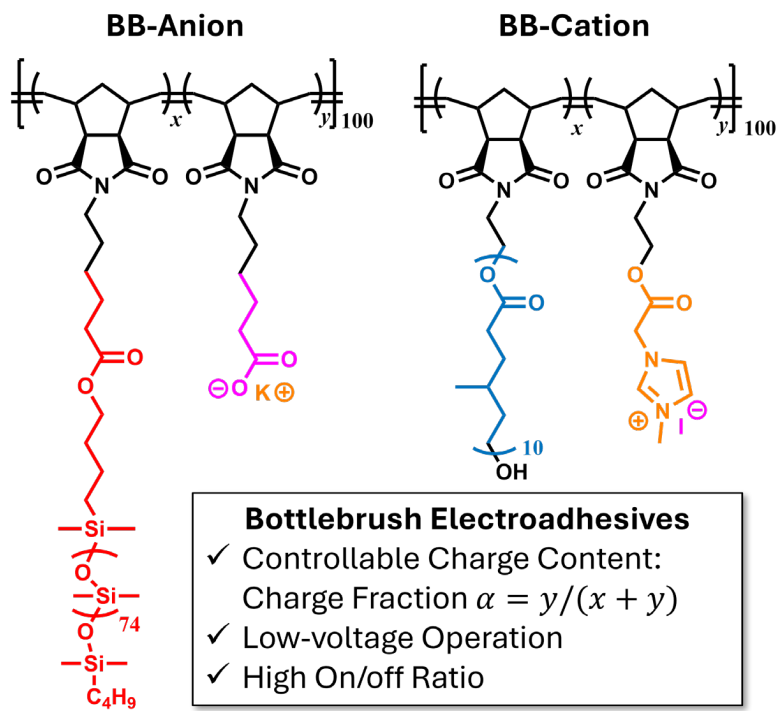

**Fig. 1| Molecular structure of pressure-sensitive electroadhesives based on bottlebrush copolymers.** Chemical structure of BB-Anion and BB-Cation bearing pendant anions and cations balanced by corresponding counterions along the backbone.

The as-synthesized BB-Anion and BB-Cation bottlebrush copolymers were formulated with bis-benzophenone-based crosslinkers, blade-coated onto a substrate, and cured under 365 nm ultraviolet (UV) light (0.30 mW/cm$^2$) for 20 minutes as shown in Fig. 2a,c and described further in Supplementary Note 1, resulting in soft, ion-containing bottlebrush elastomers with mechanical properties that resemble traditional pressure-sensitive adhesives. After blade-coating and UV-curing, smooth and defect-free films were obtained as shown in Fig. 2b, Supporting Video 1 and Fig. S29b and c, demonstrating efficient processability and compatibility with common substrates. The conformability of these bottlebrush elastomer-based adhesives was further demonstrated by contacting a BB-Cation film ($\alpha$ = 9%) prepared on a glass slide with a British one-pound coin. Optical microscope images collected through the glass slide of the coin before (Fig. 2d, left) and after contact with the polymer film (Fig. 2d, right) indicated clean, conformal contact based on differences in color without sacrificing the ability to distinguish fine embossments on the coin. No obvious voids, wrinkles or defects were observed after adhesion. Collectively, these data demonstrate that ionic bottlebrush elastomers exhibit simple processability, high-quality surfaces, and good conformability, making them promising candidates for adhesive applications.

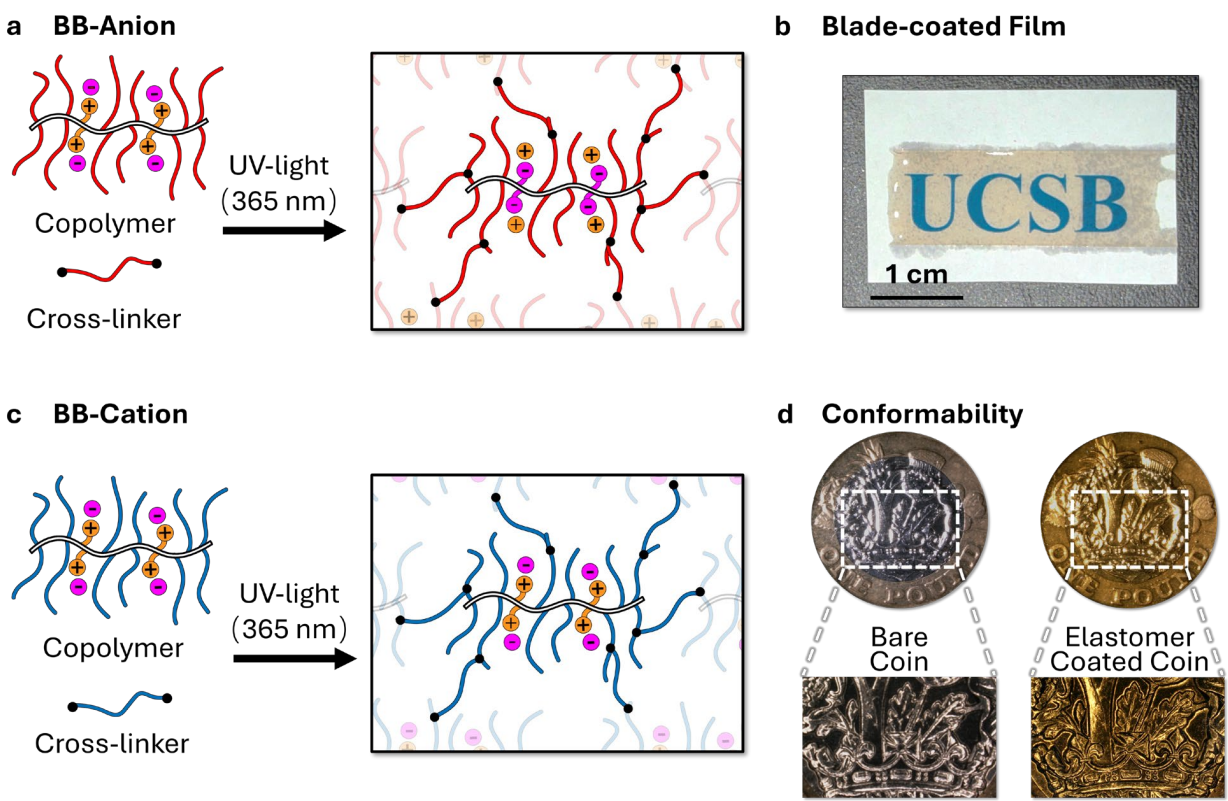

Fig. 2| **Efficient formulation, crosslinking, and processing of pressure-sensitive bottlebrush electroadhesives.** **a,c** Formulations of BB-Anion or BB-Cation efficiently crosslink under 365 nm UV light. **b**, Photograph of a crosslinked BB-Anion elastomer blade-coated on white paper demonstrates a smooth, defect-free surface morphology. **d**, Optical microscope images collected of a one-pound British coin before and after contact with a conformal BB-Cation elastomer film ($\alpha$ = 9%).

**Ion-containing bottlebrush elastomers as pressure sensitive adhesives**

The generally accepted molecular design criteria for pressure sensitive adhesives are a $T_g$ below room temperature and balanced viscoelastic dissipation in an application-specific frequency–temperature window[28]. In bottlebrush elastomer systems, thermal properties are dominated by the choice of side-chain chemistry, which constitutes the majority of each molecule by both mass and volume[29-31]. Differential scanning calorimetry was used to characterize the thermal properties of our cross-linked ion-containing bottlebrush elastomers. Figure 3a shows heat-flow traces for cross-linked BB-Cation at various charge fractions (see Supplementary Note 2 for analogous characterization of cross-linked BB-Anion). In all cases, the $T_g$ value remains essentially unchanged at approximately −53 °C, which is far below room temperature and consistent with the previously reported value for linear P4MCL[32]. The insensitivity of $T_g$ to charge fraction implies that the side chains dominate the thermal behavior of these bottlebrush elastomers.

We next performed small-amplitude oscillatory shear rheology to quantify the viscoelasticity of cross-linked BB-Cation (Fig. 3b) and BB-Anion (Fig. S18a) to examine the impact of charge fraction. In general, the storage shear modulus of BB-Cation with all four charge fractions falls in the range of 10 – 100 kPa. As expected, the bottlebrush elastomers become stiffer with increasing charge fraction, which is likely related to intermolecular crosslinking arising from ionic bonds bridged by free counterions within the network. Measured values of tan $\delta$ for BB-Cation and BB-Anion with different charge fractions are shown in Fig. S17 and S18b. Overall, tan $\delta$ values of BB-Cation and BB-Anion are consistent with previous reports[30], but decrease slightly with increasing charge fraction, again indicating a somewhat stiffer response. Notably, despite this moderate ion-induced stiffening, the crosslinked elastomers remain softer than previous polymeric electroadhesives—a consequence of the bottlebrush architecture.

To further evaluate the pressure-sensitive adhesive performance of our ion-containing bottlebrush elastomers, we mapped the rheological properties of crosslinked BB-Cation in Fig. 3c onto the viscoelastic framework proposed by Chang[33] that classifies pressure-sensitive adhesives by their storage and loss moduli ($G'$, $G''$) measured over a frequency range of $f$ = 0.01–100 Hz (see Fig. S19 for BB-Anion). Different combinations of $G'$ and $G''$ are ideal for general-purpose

PSAs (Region I, Fig. 3c), high-shear PSAs (Region II, Fig. 3c), cold-temperature PSAs (Region III, Fig. 3c), and removable PSAs (Region IV, Fig. 3c). For all charge fractions, the $G'$ values of BB-Cation remain below the Dahlquist criterion for efficient contact ($G' < 0.3$ MPa), consistent with good tack. Moreover, nearly all data points fall within Regions I–III (Fig. 3c), indicating that our bottlebrush elastomers possess suitable viscoelasticity for most PSA applications. With increasing charge fraction, the data shift toward the upper-right quadrant of the window, consistent with the aforementioned charge-induced stiffening observed in Fig. 3b.

Stress–strain curves of crosslinked BB-Cation with different charge fractions are shown in Fig. 3d. The initial slope at low strain increases with $\alpha$ and the Young's modulus of BB-Cation with 9%, 17%, and 24% charge is 42, 79, and 107 kPa, respectively (Fig. S20). The apparent fracture toughness, reflected by the area under the stress–strain curve and the strain at break, also increases with $\alpha$. For $\alpha \geqslant 17\%$, strain-softening behavior is evident, which indicates that both covalent and ionic bonds break under deformation[34]. Notably, at a charge fraction of 24%, the fracture strain exceeds 200%, a higher value than reported examples for crosslinked linear polymeric ionic liquids with much higher charge fraction (80%) used in electroadhesive applications[13], suggesting that the bottlebrush architecture is beneficial for enhancing material toughness. Collectively, these data demonstrate that ionic bottlebrush elastomers exhibit low glass-transition temperatures, softness, appropriate viscoelasticity, and toughness, together yielding mechanical properties characteristic of high-quality PSAs and making them promising candidates for pressure-sensitive adhesive applications.

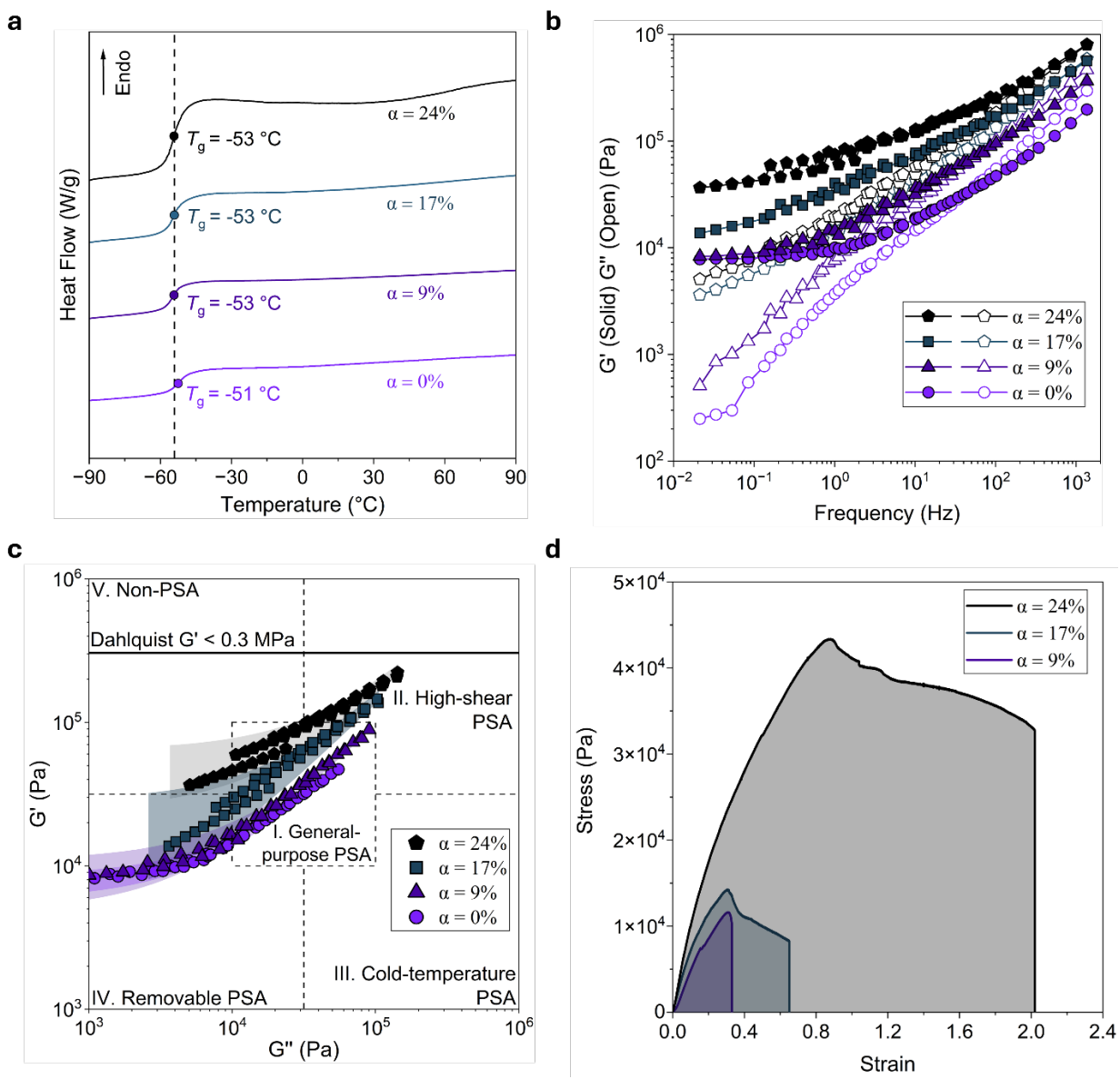

**Fig. 3| Mechanical characterization of BB-Cation bottlebrush elastomers. a,** Differential scanning calorimetry indicates charge fraction $\alpha$ has a minimal impact on the glass-transition temperature. **b,** The modulus of bottlebrush-based electroadhesives[26, 35] is significantly softer than polymeric ionic liquids with minimal stiffening observed as $\alpha$ increases. **c,** Viscoelastic window at different charge fractions over the range of $\omega \approx 0.01–100$ rad s$^{-1}$ indicates the broad applicability of bottlebrush elastomers as pressure-sensitive electroadhesives; colored backgrounds indicate a 95% confidence interval. **d,** Uniaxial tensile tests at a crosshead speed of 0.01 mm s$^{-1}$ revealed both fracture toughness and strain at break increase with $\alpha$.

**Ion-conduction and formation of ion–elastomer heterojunctions**

Electrochemical impedance spectroscopy (EIS) was performed on BB-Cation and BB-Anion to quantify ion transport and conductivity. Figure S21–S24 show EIS results for BB-Cation with different charge fractions (see also Supplementary Note 5 for BB-Anion). The data were fit to a previously reported equivalent circuit model[36]. Briefly, the model (inset in Fig. S21a) comprises two resistors corresponding to elastomer–electrode interface resistance ($R_1$) and bulk ionic transport through the elastomer ($R_2$), plus two imperfect capacitances modeled as constant-phase elements (CPEs): $Q_1$ captures the high-frequency polarization of the ion-containing network and $Q_2$ represents the electric double layer at the electrode–elastomer interface. Fitting results for all EIS measurements are summarized in Tables S2–S3. Based on the extracted value of $R_2$, the ion conductivity $\sigma$ was extracted as $\sigma = t/(R_2 A)$, where $t$ is the elastomer thickness and $A$ is the contact area with the ITO-coated electrode. The resulting $\sigma$ values for BB-Cation and BB-Anion with different charge fractions are compiled in Fig. S25 and S28, respectively. Generally, $\sigma$ falls in the range of $10^{-9}$ to $10^{-12}$ S/cm for all samples, which is low but sufficient to form a robust ionoelastomer heterojunction which is demonstrated in the following.

In the absence of an applied voltage (Fig. 4a), backbone-tethered ionic groups in ionoelastomers are locally neutralized by mobile counterions[13]. Under a favorable bias (Fig. 4b), these counterions migrate toward the electrodes, exposing oppositely charged tethered groups at the interface, forming an ionoelastomer heterojunction and strengthening adhesion through electrostatic attraction between the two polyelectrolytes[13]. To characterize the heterojunction, EIS measurements were performed on a BB-Cation/BB-Anion interface under applied direct current (DC) bias. A substantially different response was observed compared to BB-Cation or BB-Anion alone. These data were fit to an equivalent circuit as shown in Fig. 4c. Briefly, the heterojunction was modeled by an additional parallel $R_3$–$Q_3$ branch in series with the equivalent circuit used for a single bottlebrush layer (inset in Fig. S21a), where $R_3$ represents the resistance due to the interfacial ionic current and $Q_3$ is a constant-phase element to describe the double-layer capacitor at the BB-Cation/BB-Anion interface. When a favorable bias (applied DC voltage > 0 V) was applied, the interfacial resistance $R_3$ increased markedly (Table S4, Supporting Information), resulting in a large impedance in the low-frequency AC regime (<10 Hz, indicated by black arrow in Fig. 4c). This behavior confirms charging of the interfacial double layer as represented by the constant-phase element ($Q_3$) in the equivalent circuit model.

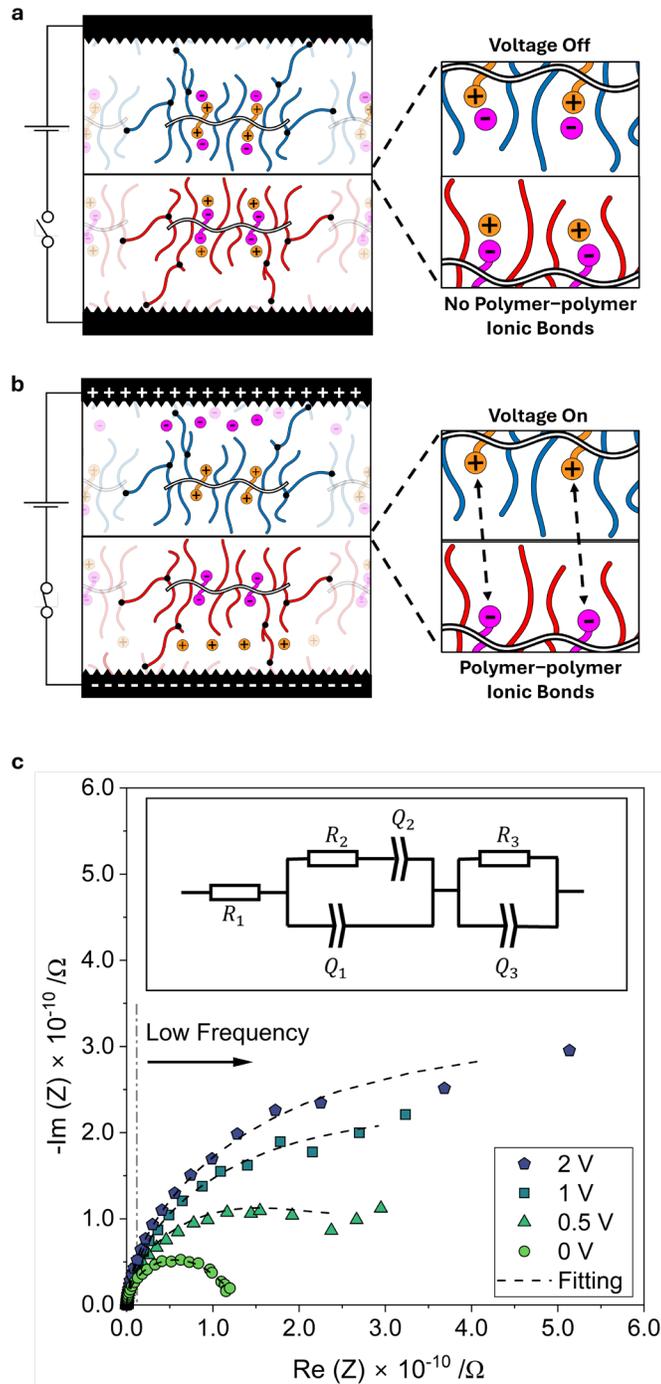

**Fig. 4| Ionoelastomer heterojunction formed by crosslinked BB-Cation and BB-Anion. a,b** Illustration of the heterojunction before (**a**) and after (**b**) an external voltage is applied. **c,** Nyquist impedance plot of an ion–elastomer heterojunction formed from BB-Cation ($\alpha = 17\%$) and BB-Anion ($\alpha = 51\%$) under different DC biases. Dashed lines represent fits to the equivalent circuit model. The data points were collected at different frequencies, ranging from 3 MHz for the leftmost point to 0.05 Hz for the rightmost point. The low-frequency region (0.05−10 Hz, indicated by the black arrow) and high-frequency region (10−3000000 Hz) are separated by a vertical gray dashed-dotted line.

**Electroadhesion of ion-containing bottlebrush elastomers**

Indentation–adhesion tests were conducted using a custom-built JKR test frame to evaluate the adhesive performance of ion-containing bottlebrush elastomers. Samples were fabricated by blade coating formulations containing as-synthesized bottlebrush polymer, crosslinker, and photoinitiator onto a porous carbon electrode (Fig. 5a, left and middle; Supplementary Video 2). The resulting uniform and smooth film (Fig. S29a) was cured under 365-nm light for 20 minutes (Fig. 5a, right), followed by lamination with a conductive polyimide layer and mounting on a custom-built adhesion fixture (Fig. 5b). Adhesion was measured using a cross-cylinder geometry (Fig. 5b and c), which is equivalent to a sphere-on-flat contact under the Derjaguin approximation[13, 37].

The adhesion test comprises three stages—approach, dwell, and separation—with representative force–displacement traces shown in Fig. 5c and Fig. S30. The fixtures holding a complementary pair of ion-containing elastomers were initially separated by 1 mm. The upper fixture was then lowered at 0.01 mm s$^{-1}$ until contact was established (*approach*, Fig. S30). The samples were held in contact for 300 s under a constant normal load of 200 mN as regulated by a proportional–integral–derivative (PID) controller (*dwelling*, Fig. S30) to allow interfacial equilibration (the dwell time was determined by a dynamic study; see Supplementary Note 9 for details). Finally, the upper fixture was raised at the same rate (0.01 mm s$^{-1}$) back to its original position (*separation*, Fig. S30). A sharp debonding peak typically appeared at some point in the separation stage, from which the work of adhesion was obtained as the area under the negative force–displacement curve. Figure 5d shows the voltage-dependent adhesion between BB-Cation ($\alpha = 17\%$) and BB-Anion ($\alpha = 51\%$). Notably, the adhesion force increased as the external voltage was raised, and under a favorable bias of 5 V the adhesion force is nearly tripled relative to the intrinsic adhesion at 0 V. Supporting Video 3 and 4 further demonstrate electroadhesion of the 17% BB-Cation / 51% BB-Anion pair under an applied bias of just 2 V: when the voltage was turned on, the adhesive bears a load (Supporting Video 3), while turning off the voltage caused deadhesion within ~1.2 s (Supporting Video 4). Notably, BB-Cation ($\alpha$ = 17%) also forms an electroadhesive heterojunction with a linear counterpart, poly[1-(2-acryloyloxyethyl)-3-butylimidazolium] bis(trifluoromethane)sulfonimide (Figure S32).

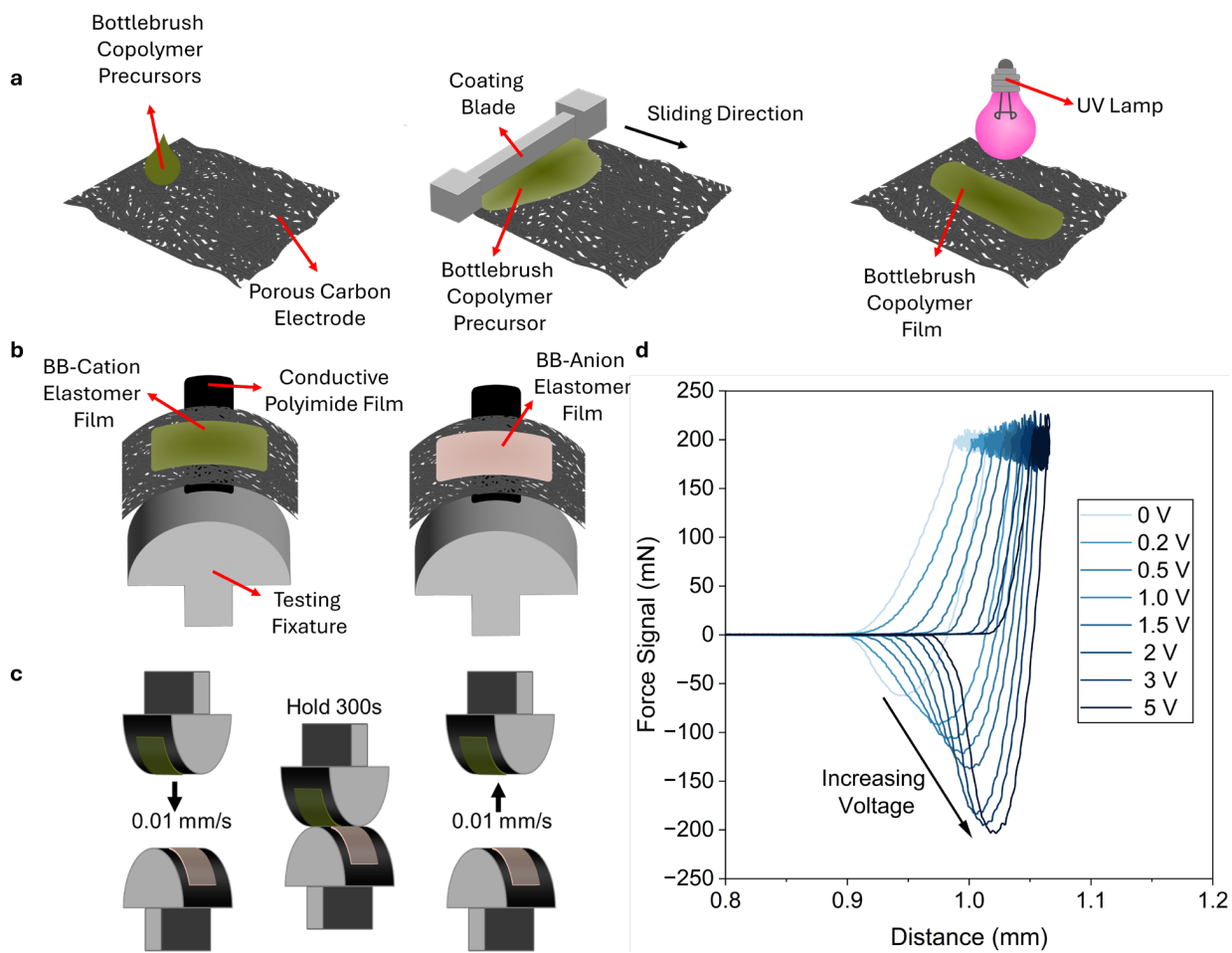

**Fig. 5| Low-voltage response of ion-containing bottlebrush electroadhesives. a,** Crosslinked BB-Cation and BB-Anion elastomers were blade-coated on porous carbon electrodes. **b**, Films were glued onto a half-cylinder fixture for adhesion testing. **c,** Adhesion tests were performed at a constant approach speed (0.01 mm/s) with a dwell time of 300 s and a constant separation speed (0.01 mm/s). The force was monitored and recorded throughout all stages. **d,** JKR curves collected from measurements of BB-Cation ($\alpha$ = 17%) in contact with BB-Anion ($\alpha$ = 51%) at different applied voltages.

Adhesion measurements were similarly performed for heterojunctions formed between BB-Cation and BB-Anion having different charge fractions, as well as between BB-Cation and the aforementioned linear anionic polymer. The intrinsic adhesion at 0 V for all three heterojunctions increases with increasing charge fraction in BB-Cation (see Supplementary Note 8 for details). To exclude the contribution of intrinsic adhesion from the analysis of electrically induced adhesion, the adhesion values were normalized by the corresponding intrinsic adhesion measured in the absence of an applied electric field, as summarized in Fig. 6a and Fig. S34a,b. Normalized adhesion, which represents the ratio of adhesive strength when the voltage is turned on vs. off (the "on/off ratio"), increased with charge fraction as expected. Larger charge fractions lead to greater ion migration to the electrodes and a higher density of exposed ionic groups at the interface, which strengthens polymer–polymer electrostatic interactions. The results are also consistent with a previous model in which adhesion is proportional to the interfacial double-layer capacitance, $Q_3$ (Table S4)[38]. It should be noted that film quality is important for accurate adhesion measurements and any presence of defects such as air bubbles after blade coating results in an artificially lower on/off ratio (see Supplementary Note 8 and 10 for details). To improve the consistency of film coating and adhesion measurements, a treatment step was introduced for the 24% charge fraction BB-Cation in which the BB-Cation film was annealed at 80 ℃ under vacuum for 24 h after blade-coating (but before crosslinking) to improve surface uniformity. This post-coating treatment was found to increase the maximum on/off ratio (voltage = 5 V) from 1.63 to 5.11 for the 24% BB-Cation/51% BB-Anion pair. As discussed in Supplementary Note 11, calculations of the charge density (charge per unit mass) for BB-Cation and BB-Anion samples indicate they are approximately matched for the 24% BB-Cation/51% BB-Anion pair, and for which the highest on/off ratio was achieved, suggesting that charge density matching may be an important design parameter for ion–elastomer heterojunctions.

The on/off ratio also increases with external voltage (Fig. 5d and 6a) for favorable bias voltages in the range of 0 to 2 V. Above 2 V, the on/off ratio reaches a plateau across all adhesive pairings (Fig. 6a). This indicates that ion-containing bottlebrush elastomers can be effectively operated at 2 V, which is lower than the operating voltage of dielectric-based electroadhesives[22], and this on–off ratio exceeds values previously reported for poly(ionic liquid)-based systems[13, 27, 38] (Fig. 6b). The switchability of bottlebrush electroadhesives is further demonstrated in Fig. 6c and Supporting Video 4. When a voltage was applied across the heterojunction, an increase in adhesive force supports a substantial stress (343 mN/cm$^2$). Upon turning off the voltage, the weight drops within ~1.2 s, clearly demonstrating rapid switchability.

Intriguingly, the charge density of the ionic bottlebrush polymers is substantially lower than that of linear poly(ionic liquid)-based electroadhesive materials. For BB-Cation with $\alpha = 24\%$ (excluding counterions), the total charge density is ~18 C/g (Table S5, charge concentration is estimated as 18 C/cm$^3$, see Supplementary Note 11 for details), compared to ~346 C/g for the aforementioned linear poly(ionic liquid) (excluding counterions, Table S7)[13]. This approximately order of magnitude difference suggests that either linear poly(ionic liquids) possess many more charges than are necessary for efficient electroadhesion, or that the bottlebrush architecture with long, flexible side chains significantly markedly lowers the charge density needed for efficient electroadhesion (18 vs. 346 C/g), even though the charge fractions are of the same order (24% vs. 80%). This observation opens avenues for further investigation of ion diffusion effects, including how ion concentration distributes across ion-elastomer heterojunctions with different molecular architectures and how these features influence electroadhesive strength.

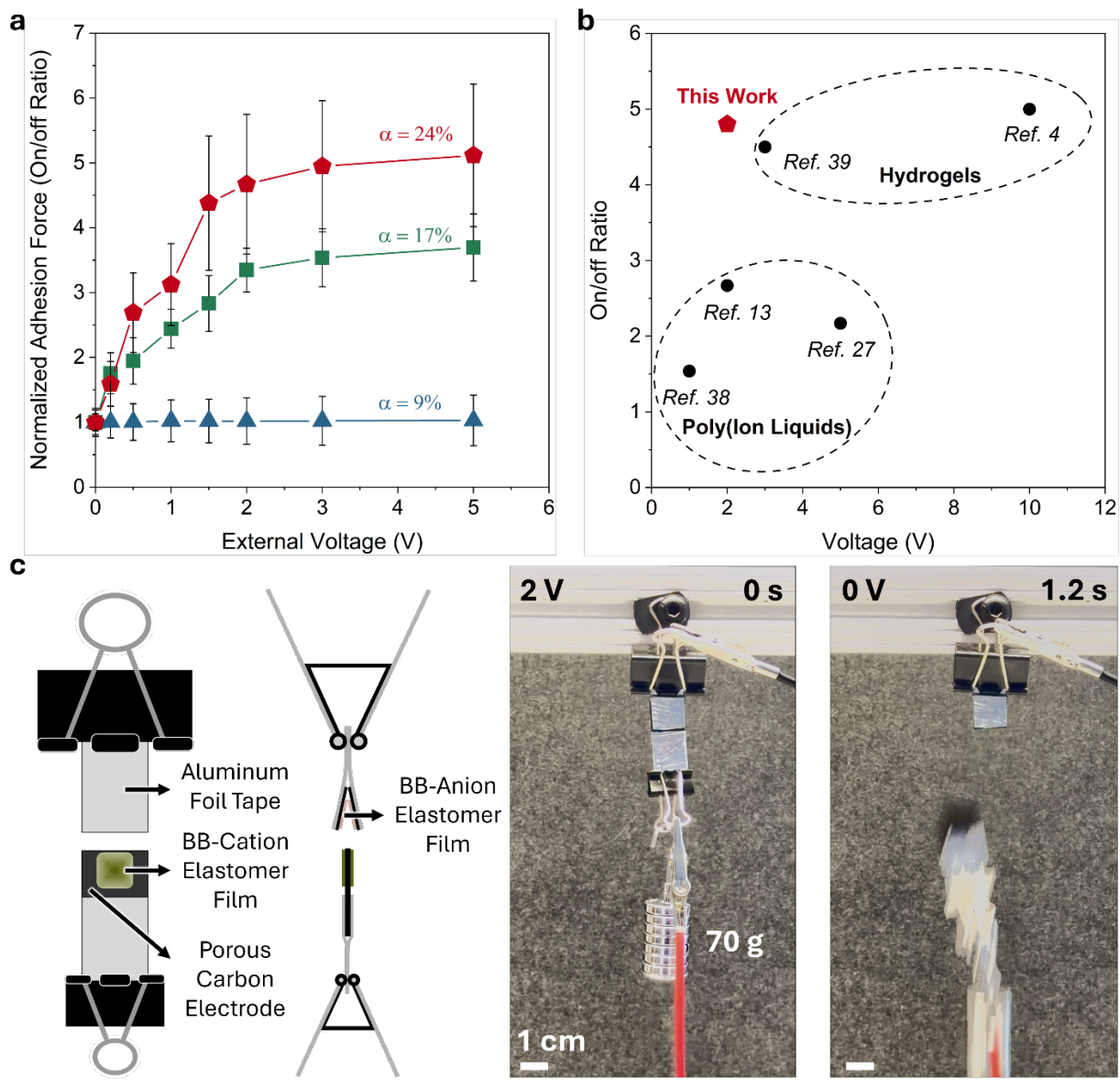

**Fig. 6| Electroadhesion measurements and switchability of BB-Cation and BB-Anion. a,** Normalized electroadhesion of BB-Cation with different charge fractions against BB-Anion (α = 51%). **b,** Comparison with performance of low-voltage (< 10 V) electroadhesives previously reported in the literature. **c,** Switchable adhesion between BB-Cation and BB-Anion. *Left*: Illustration of the sample configuration. *Middle*: Photograph of the sample hanging with an applied load under an external voltage of 2 V. *Right*: Photograph showing the load dropping within ~1 s after the voltage is switched off; the rapid dropping motion creates a blur within the image.

## Conclusions

This study demonstrates that ion-containing bottlebrush elastomers are a promising platform for the design of pressure-sensitive electroadhesives. Bottlebrush polymer precursors bearing independently controllable amounts of anionic or cationic charges were readily synthesized by ring-opening metathesis polymerization, followed by formulation with bis-benzophenone crosslinkers and UV curing, yielding materials with substantially reduced entanglement as well as tunable viscoelasticity and mechanical properties in a range characteristic of PSAs. Ionic bottlebrush electroadhesives operate efficiently at ion densities as low as 18 C/g—one order of magnitude lower than linear poly(ionic liquids), despite having charge fractions of the same order. By operating at low voltages (~2 V) with high on/off ratios exceeding 4.5, these materials combine the performance of poly(ionic liquid) heterojunctions and switchable hydrogels in an all-solids formulation. The insights developed through this work create opportunities to expand the scope of polymeric electroadhesives to applications that would benefit from fast and reversible adhesion with conformable pressure-sensitive materials.

## Materials and Methods

**Materials.** *N*-(3-Dimethylaminopropyl)-*N′*-ethylcarbodiimide hydrochloride (EDC·HCl, Sigma Aldrich, commercial grade), 4-(Dimethylamino)pyridine (DMAP, Sigma Aldrich), Chloroform (Sigma Aldrich, anhydrous, contains amylenes as a stabilizer, purity ⩾ 99%), acetone (Fisher Scientific, purity ⩾ 99.5%), methanol (Fisher Scientific, purity = 99.5%), dichloromethane (Fisher Scientific, purity ⩾ 99.5%), ethyl vinyl ether (ACROS Organic, Contains 0.1% KOH as stabilizer, purity = 99%), dimethylformamide (dry over molecular sieve), CDCl$_3$ (Cambridge Isotope Laboratories, purity = 99.8%), DMSO-*d$_6$* (Cambridge Isotope Laboratories, purity = 99.9%), and acetone-*d$_6$* (Cambridge Isotope Laboratories, purity = 99.9%), KOH (Sigma Aldrich), tetrahydrofuran (THF, Fisher Scientific) were used as received. *N*-(hydroxyethyl)-*cis*-5-norbornene-*exo*-2,3-dicarboximide, Grubbs' third-generation metathesis catalyst [(H₂IMes)(pyr)₂(Cl)₂Ru=CHPh] (G3), *N*-(hydroxyethyl)-cis-5-norbornene-exo-2,3-dicarboximide (Nb-OH), Poly-4-methylcaprolactone (P4MCL), 1*H*-Imidazol-1-ylacetic acid (Sigma Aldrich). PMCL macromonomer (Nb-P4MCL), N-(hexanoic acid)-cis-5- norbornene-*exo*-dicarboximide (Nb-COOH) and Poly-dimethylsiloxane (PDMS) macromonomer (Nb-PDMS) were synthesized in-house (see Supplementary Note 1 for details).

**Preparation of ion-containing elastomers.** Poly[norbornene-poly(4-methylcaprolactone) (Nb-P4MCL)]-b-poly[norbornene-imidazolium]+I− (BB-Cation) and poly[norbornene-polydimethylsiloxane (Nb-PDMS)]-b-poly[norbornene-carboxylate]−K+ (BB-Anion) were synthesized via ring-opening metathesis polymerization. The as-synthesized polymers were cross-linked under UV-light (wavelength: 365 nm, power: 0.3 mW/cm²). See Supplementary Note 1 for details.

**Conformability Test.** The BB-Cation copolymer ($\alpha$ = 9%) formulated with cross-linker was blade-coated onto a glass slide and subsequently UV-cured for 20 min at 365 nm with an intensity of 0.3 mW/cm². The resulting glass-supported cross-linked BB-Cation elastomer was then brought into contact with a one-pound British coin. Images of the coin in contact with the elastomer film were collected through the glass slide using a Keyence VHX-5000 microscope. For comparison, images of the bare coin covered with a bare glass slide were also acquired in the same way.

**Differential scanning calorimetry (DSC).** Approximately 5 mg of cross-linked, ion-containing bottlebrush elastomer was placed in a Tzero aluminum pan and hermetically sealed. DSC was then performed on a TA Instrument DSC 2500s from −100 to 100 °C at 10 °C min⁻¹ with 5-min isotherms at each stop temperature and a cooling rate of 10 °C min⁻¹. Data reported here and glass-transition temperatures ($T_g$) were collected and determined from the second heating cycle.

**Rheology.** A strain-controlled ARES-G2 rheometer (TA Instruments) equipped with a liquid nitrogen dewar was used to characterize shear stress relaxation and linear viscoelastic properties of the cross-linked bottlebrush elastomers. All measurements employed 8 mm stainless steel parallel plates. Strain-sweep experiments were first conducted to determine the linear viscoelastic region (LVE). For time–temperature superposition (TTS), isothermal frequency sweeps from 100 to 0.1 rad s$^{-1}$ at a constant strain of 1% were obtained at multiple temperatures, and master curves were constructed using the Williams–Landel–Ferry (WLF) relation.

**Uniaxial tensile tests.** The as-synthesized BB-Cation was mixed with a cross-linker at a molar ratio of repeat units to cross-linker of 1:15. The mixture was blade-coated onto a silicone wafer pre-treated with a mold-release spray (Mitreapel BK30). The as-cast film (20 mm × 5 mm × 1 mm, length × width × thickness) was then UV-cured (365 nm) for 20 min to induce cross-linking. The resulting cross-linked film was mounted on a texture analyzer (Stable Micro Systems TA.XTPlus Connect located within the BioPACIFIC Materials Innovation Platform at UC Santa Barbara), and uniaxial tensile loading was applied at a crosshead speed of 0.01 mm s$^{-1}$ until failure.

**Sample fabrication for adhesion tests.** Samples for adhesion tests were prepared by blade coating. Two 500-μm-thick spacers were affixed to a porous carbon electrode (Sigracet 22 BB; Fuel Cell Store) with a 1 cm horizontal separation, and ~300 mg uncross-linked ion-containing bottlebrush precursor was dispensed between the spacers (Supplementary Video 1, 2). A blade was used to draw the liquid-like precursor and fill the gap. The wet film was then degassed under vacuum at 60 °C for 1 h. For the 24% charge-fraction BB-Cation, the as-coated film was additionally annealed at 80 °C under vacuum for 24 h to improve surface quality. After degassing (and annealing if any), films were UV-cross-linked for 20 min. The resulting films were roughly 1 cm wide, 2 cm long, and 0.5 mm thick. For BB-Anion, the resulted films with substrates were soaked in 0.05 M KOH water/THF solution for 24 h to dope. After doping, the film was dried in the air for 24 h.

**Electrochemical impedance spectroscopy (EIS).** Impedance measurements on neat ion-containing bottlebrush elastomers were conducted in a parallel-plate cell formed by indium tin oxide (ITO)–coated glass electrodes. The precursor was cast onto one ITO slide within a circular aperture defined by Kapton tape (the radius of the circle is 0.318 mm for BB-Cation and 0.476 for BB-Anion samples) and covered with a second ITO slide. The measurements were performed using Biologic's Intermediate Temperature System (ITS) in conjunction with a VSP-300 potentiostat. A sinusoidal voltage with an amplitude of 100 mV for BB-Cation or 300 mV for BB-Anion was applied in the frequency range of 0.1 Hz − 3 MHz. Data were then fit to the equivalent circuit shown in Fig. S21a inset to extract the resistance. From these equivalent direct-current (DC) resistances, the conductivity was calculated according to $\sigma = \frac{t}{R_2 A}$, where $t$ is the elastomer thickness, $R_2$ is a component in the equivalent circuit model (Fig. S21a inset) and $A$ is the contact area with the ITO-coated electrode. The total thickness of the whole cell was measured, and the thickness of the loaded sample was obtained by subtracting the thickness of the ITO glasses from the total thickness. Similarly, impedance measurements of the ionoelastomer heterojunction were conducted in a parallel-plate cell formed by ITO-coated glass electrodes. BB-Cation and BB-Anion elastomer films were prepared separately on different ITO-coated glass slides, which were then brought into contact. The precursor of each elastomer was cast onto the corresponding ITO slide within a rectangular aperture defined by Kapton tape (1 cm × 1 mm). A sinusoidal voltage with an amplitude of 300 mV was applied over a frequency range of 0.05 Hz to 3 MHz. Data were collected at applied DC biases of 0, 0.5, 1, and 2 V, and then fitted to the equivalent circuit shown in the inset of Fig. 4c to extract the circuit parameters.

**Adhesion tests.** Cross-linked ion-containing bottlebrush elastomer films were laminated to a conductive polyimide film ($L \times W$ = 2 × 1 cm, McMaster-Carr) using conductive carbon double-sided tape. The elastomer/carbon-electrode/polyimide assembly was mounted to a custom half-cylinder fixture with double-sided tape (radius, 10 mm; design in Fig. S31). A commercial texture analyzer (Stable Micro Systems TA.XTPlus Connect, located within the BioPACIFIC Materials Innovation Platform at UC Santa Barbara) recorded normal load and displacement during contact-adhesion tests. The two elastomer-covered fixtures were installed on the instrument, and a DC power supply was connected by clamping the polyimide backings. The current power supply was limited to 1 mA to prevent possible short circuiting or leakage. In addition, the portion of the testing fixture in contact with the texture analyzer was wrapped with 3M Scotch tape, which served as an insulating material. Before contact, a voltage bias was applied with the fixtures separated by 1 mm. The upper fixture was then lowered at 0.01 mm s$^{-1}$ until the normal load reached 200 mN, followed by a 300 s dwell. Finally, the fixtures were separated at 0.01 mm s$^{-1}$ while continuously recording load–displacement.

## Acknowledgement

This work was supported by the MRSEC Program of the National Science Foundation under Award No. DMR 2308708 (IRG1) and made use of the BioPACIFIC Materials Innovation Platform supported by Awards DMR-1933487 and DMR-2445868. This work was also partially supported by NSF CMMI 2053760 (synthesis). C.M.B. also thanks The Camille and Henry Dreyfus Foundation for partial support. The authors thank Professor Claus Eisenbach for inspiring discussions and Carina Lin for helping collect demonstration photos and videos.

# Supplementary Information

# Ion-Containing Bottlebrush Elastomers as Pressure-Sensitive Electroadhesives

*Hao Dong[1,#], Intanon Lapkriengkri[1,2,#], Nadia Chapple[2], Hyunki Yeo[1,2], Alexandra Zele[2], Hiba Wakidi[1,3], Thuc-Quyen Nguyen[1,3], Michael L. Chabinyc[1,2], Christopher M. Bates[1,2,3,*], Megan T. Valentine[1,4,*]*

[1]Materials Research Laboratory, University of California Santa Barbara, Santa Barbara, CA 93106, USA.
[2]Materials Department, University of California Santa Barbara, Santa Barbara, CA 93106, USA.
[3]Department of Chemistry and Biochemistry, University of California Santa Barbara, Santa Barbara, CA 93106, USA.
[4]Department of Mechanical Engineering, University of California Santa Barbara, Santa Barbara, CA 93106, USA.

# These authors contributed equally to this work.
* Corresponding Authors

Emails:
* E-mail: cbates@ucsb.edu, valentine@engineering.ucsb.edu

## 1. Materials and ion-containing bottlebrush elastomer synthesis

**Materials synthesized in-house.** *N*-(hydroxyethyl)-*cis*-5-norbornene-*exo*-2,3-dicarboximide was prepared according to the literature[1]. Grubbs' second-generation metathesis catalyst [(H$_2$IMes)(PCy$_3$)(Cl)$_2$Ru=CHPh] was purchased from Sigma Aldrich. Grubbs' third-generation metathesis catalyst [(H$_2$IMes)(pyr)$_2$(Cl)$_2$Ru=CHPh] (G3) was prepared according to the literature[2]. *N*-(hydroxyethyl)-cis-5-norbornene-exo-2,3-dicarboximide (Nb-OH) was synthesized according to literature[3]. Poly-4-methylcaprolactone (P4MCL) macromonomer (Nb-P4MCL) was prepared using previously reported methods[3]. Tin(II) 2-ethylhexanoate (Sn(Oct)$_2$, Sigma Aldrich, 92.5–100%) was purified according to literature[4]. 1*H*-Imidazol-1-ylacetic acid (Sigma Aldrich). N-(hexanoic acid)-cis-5- norbornene-*exo*-dicarboximide (Nb-COOH) and Poly-dimethylsiloxane (PDMS) macromonomer (Nb-PDMS) was synthesized according to literature[5]. The macromonomer side-chain degree of polymerization ($N_{SC}$) was 10 for BB-Cation and 74 for BB-Anion as measured by $^1$H nuclear magnetic resonance and gel-permeation chromatography. The cross-linkers for BB-Cation, 3-methylpentane-1,5-diyl bis(4-benzoylbenzoate) (EMHbisBP) and for BB-Anion, polydimethylsiloxane bis(4-benzoylbenzoate) (PDMSbisBP) were synthesized following previous protocols[6]. Poly[1-(2-acryloyloxyethyl)-3-butylimidazolium] bis(trifluoromethane)sulfonimide (Linear-Anion) was synthesized as reported in literature[7].

**Synthesis**

*Imidazole functionalized norbornene anhydride*

In a round bottom flask charged with a stirbar, *N*-(hydroxyethyl)-cis-5-norbornene-exo-2,3-dicarboximide (Nb-OH) (10 g, 48 mmol), 1*H*-Imidazol-1-ylacetic acid (12.1 g, 96 mmol), *N*-(3-Dimethylaminopropyl)-*N'*-ethylcarbodiimide hydrochloride (EDC·HCl) (28 g, 144 mmol), and 4-(Dimethylamino)pyridine (DMAP) (2.9 g, 24 mmol) were added. 250 mL of chloroform was added and the cloudy mixture was stirred at room temperature for 48 hours. The resulting solution was repeatedly washed with water then dried with magnesium sulfate. The yellow liquid was filtered through a plug of basic alumina and activated carbon to yield a clear pale-yellow liquid. The product was dried *in vacuo* and analyzed using NMR and mass-spectrometry. Yield = 96%.

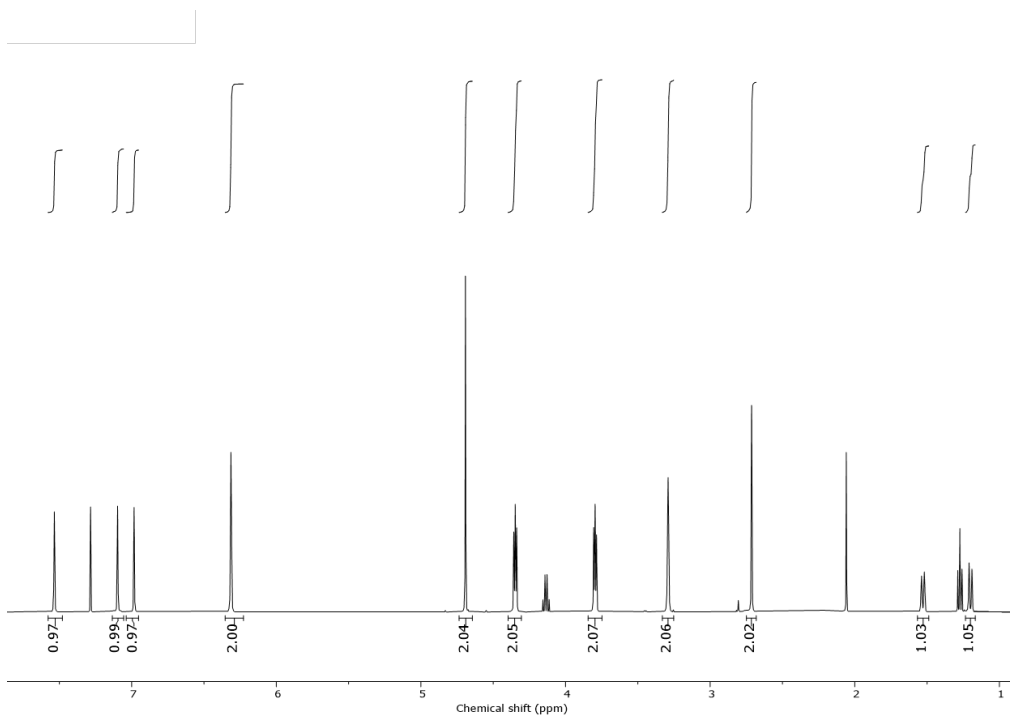

**Fig. S1.** $^1$H NMR (CDCl$_3$) of Imidazole functionalized norbornene anhydride.

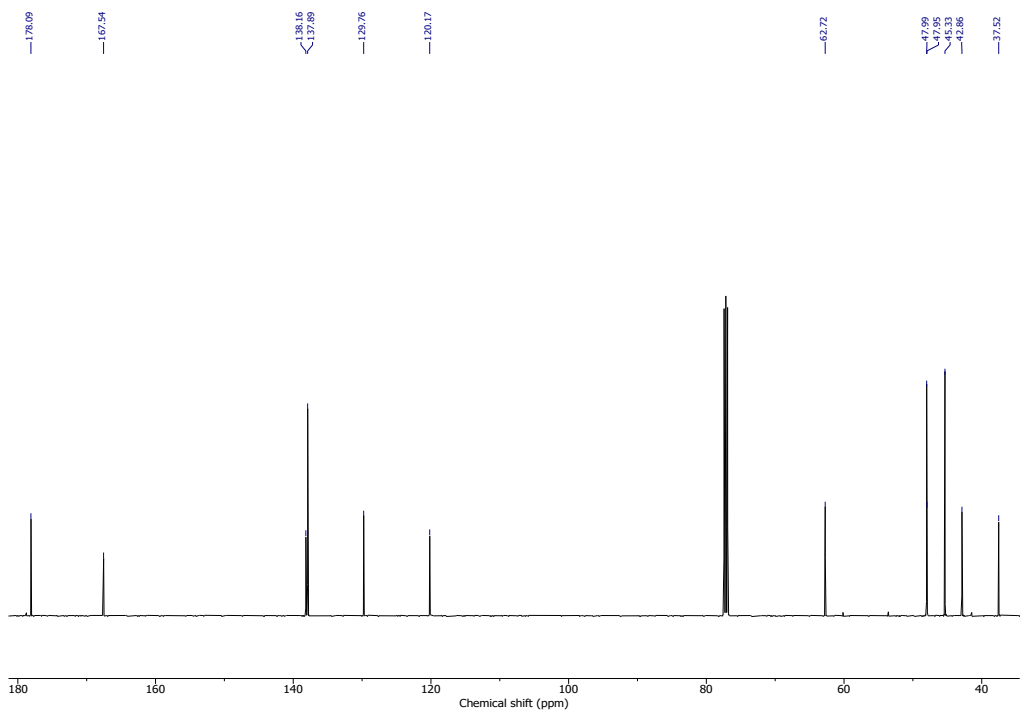

**Fig. S2.** $^{13}$C NMR (CDCl$_3$) of Imidazole functionalized norbornene anhydride.

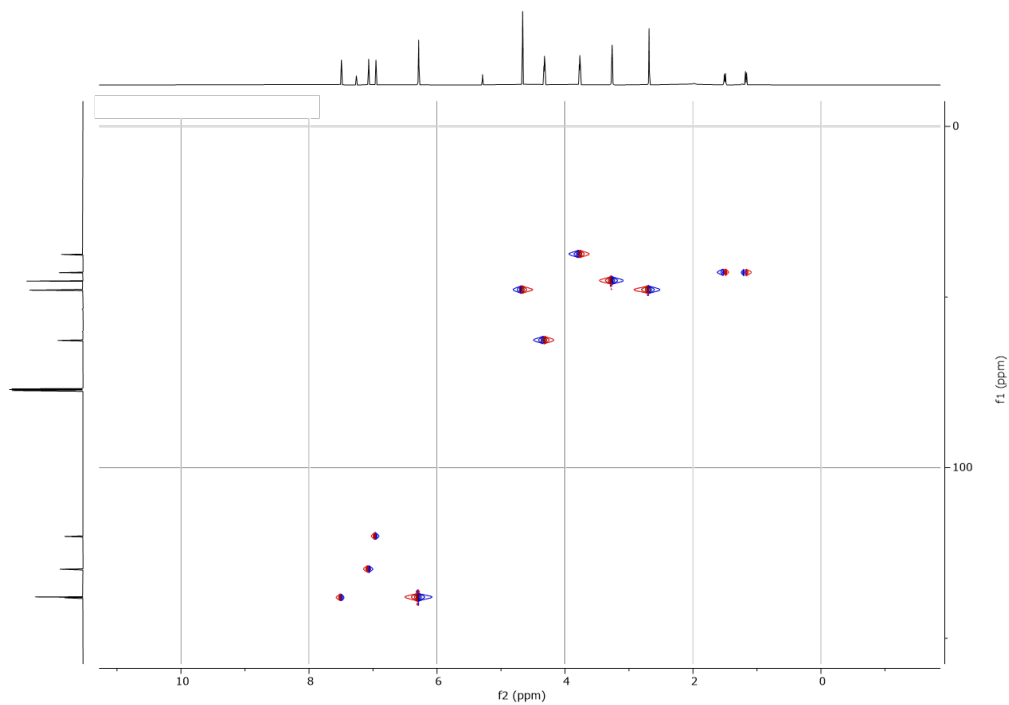

**Fig. S3.** HSQC $^1$H-$^{13}$C NMR (CDCl$_3$) of Imidazole functionalized norbornene anhydride.

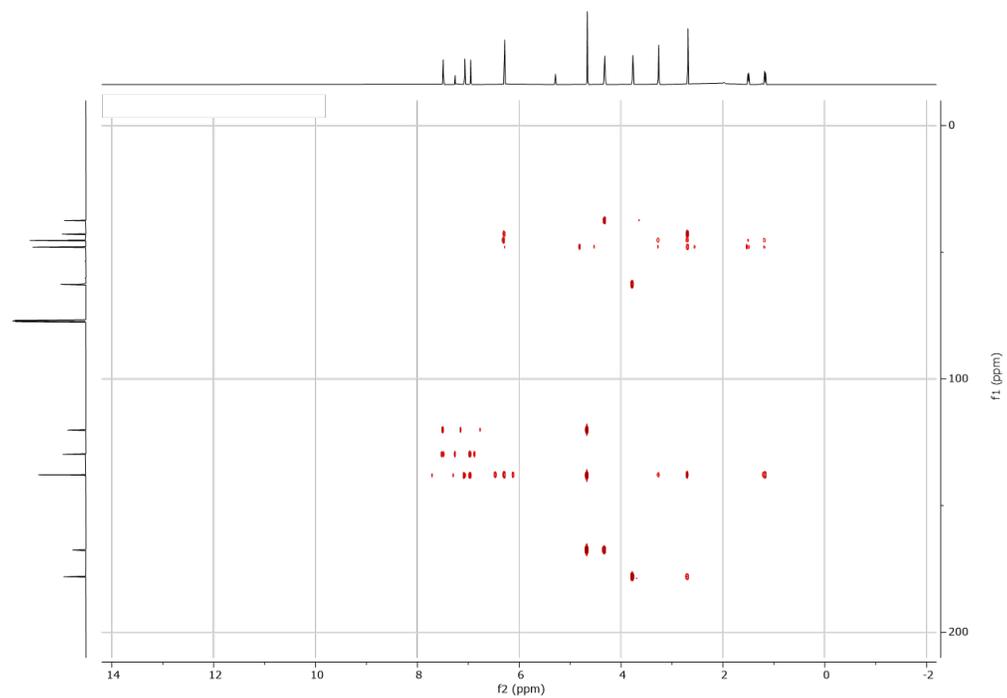

**Fig. S4.** HMBC $^1$H-$^{13}$C NMR (CDCl$_3$) of Imidazole functionalized norbornene anhydride.

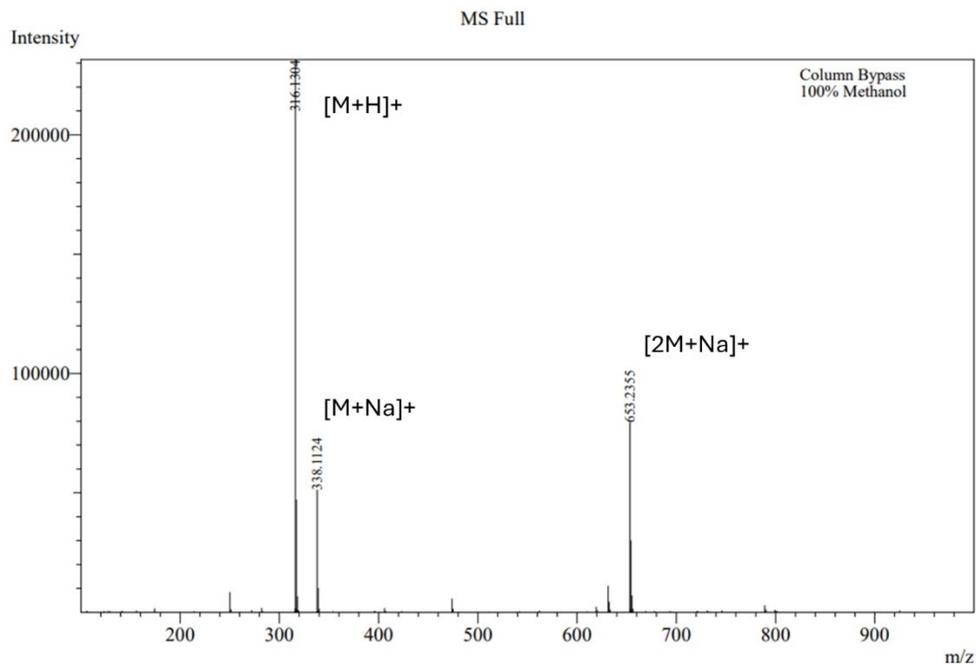

**Fig. S5.** ESI-MS scan of Imidazole functionalized norbornene anhydride.

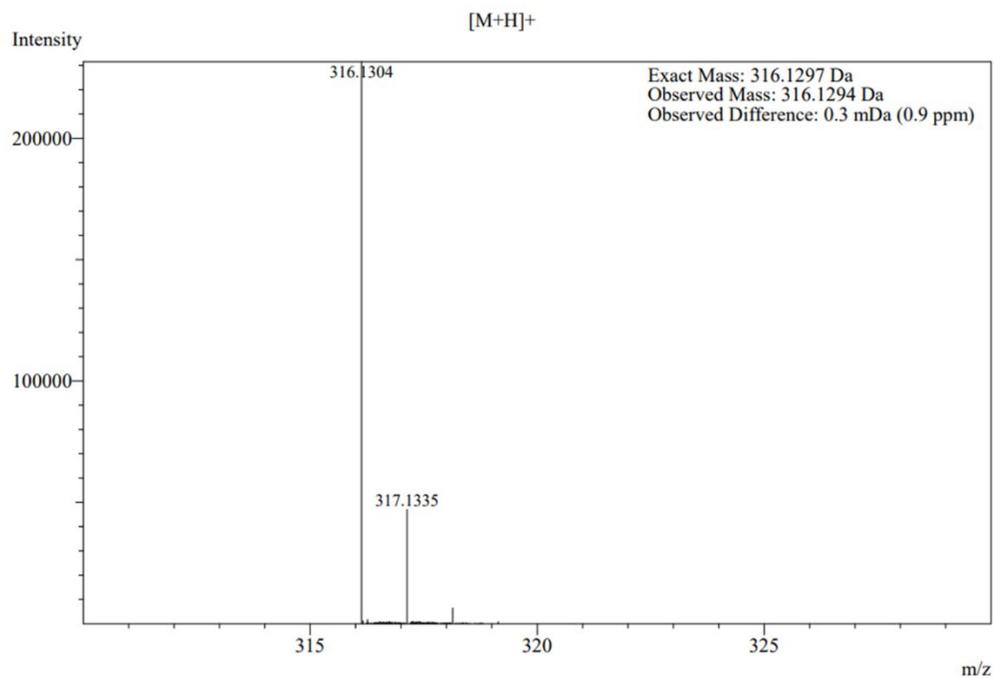

**Fig. S6.** Accurate mass scan of Imidazole functionalized norbornene anhydride.

*Imidazolium functionalized norbornene anhydride*

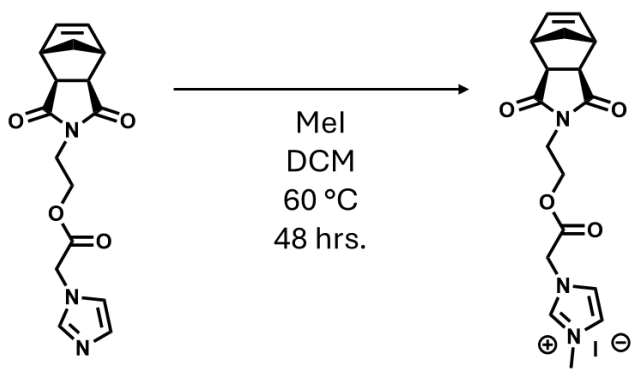

In a dried pressure flask charged with a stir bar, imidazole functionalized norbornene anhydride (6 g, 18.1 mmol), and methyl iodide (7.7 g, 54,2 mmol) were added and dissolved in 100 mL of dried DCM. The flask was sealed with a bottom-seating PTFE lined cap and a perfluoro O-ring. The solution was heated to 60 °C while stirring for 48 hours before it was allowed to cool to room temperature. The yellow solution was dried *in vacuo* yielding an orange solid. The product was analyzed with NMR and IR spectroscopy. Yield = 98%.

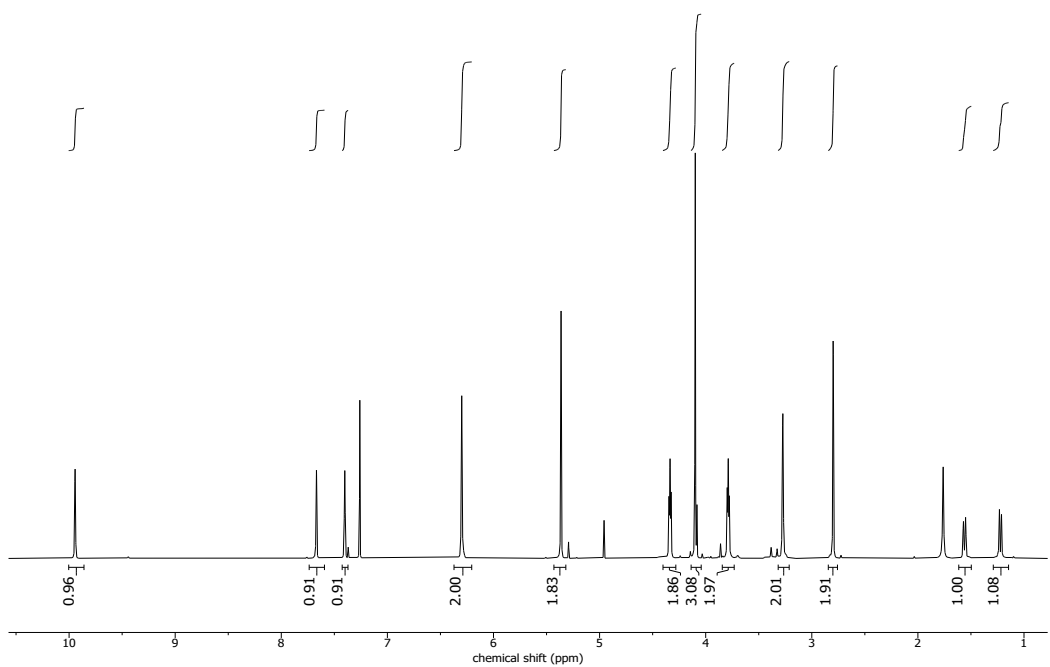

**Fig. S7.** $^1$H NMR (CDCl$_3$) of Imidazolium functionalized norbornene anhydride.

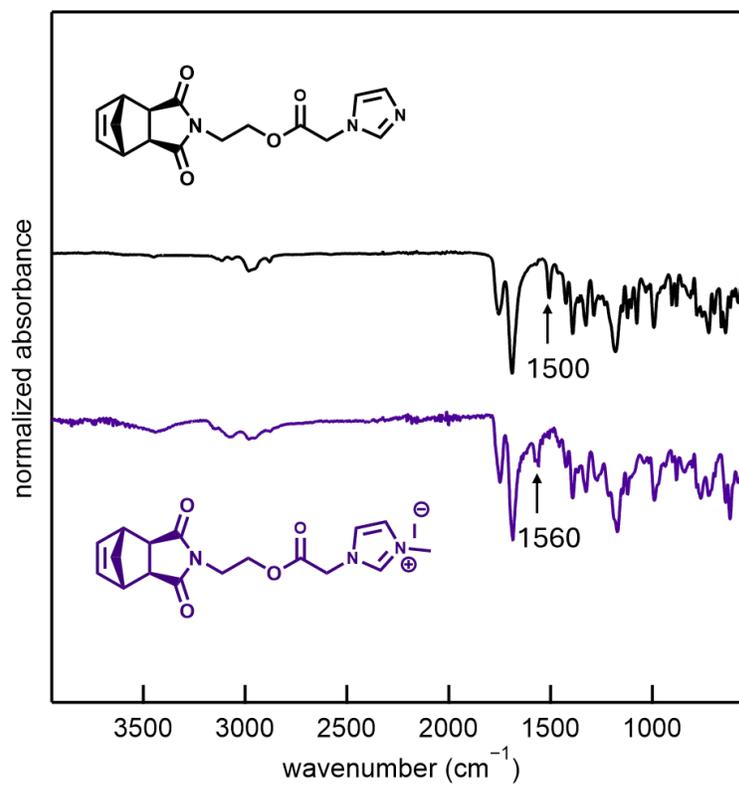

**Fig. S8.** IR spectroscopy comparison between imidazole functionalized and imidazolium functionalized norbornene anhydride.

*Imidazolium functionalized P4MCL bottlebrush (BB-Cation Copolymer)*

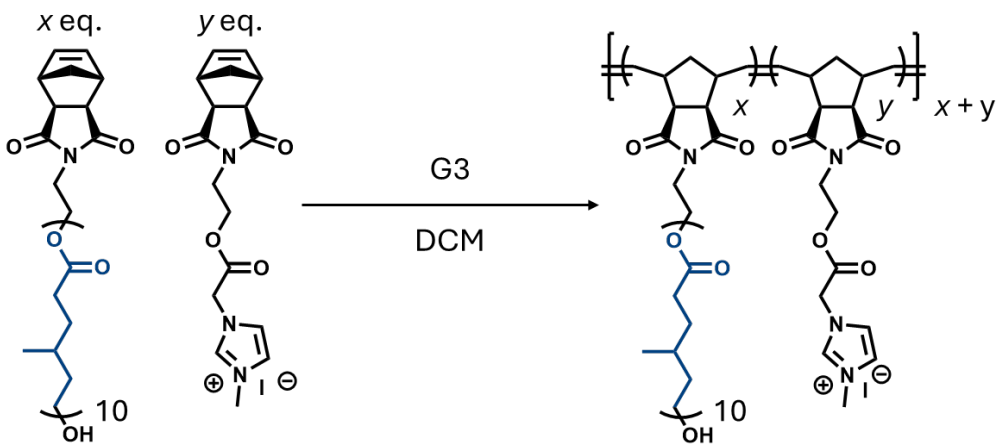

A grafting-through polymerization of P4MCL macromonomer and imidazolium functionalized norbornene anhydride monomer was performed in a diluted solution of macromonomer and monomer. The monomeric charged fraction of the resulting polymer was controlled through the added ratio of P4MCL macromonomer and imidazolium functionalized norbornene anhydride monomer. For example, a 10% nominal charged imidazolium functionalized bottlebrush (actual charge fraction: 9%) can be made as follows. In a nitrogen-filled glovebox, P4MCL macromonomer (2 g, 1.33 mmol), and imidazolium functionalized norbornene anhydride monomer (70.1 mg, 0.15 mmol) were dissolved in anhydrous DCM. In a separate container, G3 (10.71 mg, 0.015 mmol) was dissolved in 0.2 mL of anhydrous DCM before rapid injection into a vigorously stirring solution of the macromonomer and monomer. After 24 hours, the polymerization was removed from the glovebox and terminated using a large excess ethyl vinyl ether (EVE) (>100 eq) with stirring for 30 minutes. A crude $^1$H NMR of the freshly quenched reaction was taken. The disappearance of a peak at 6.3 ppm corresponding to unreacted olefin protons was observed, suggesting high reaction conversion. The resulting reaction mixture was concentrated *in vacuo* and the polymer precipitated into hexanes. The imidazolium functionalized

P4MCL bottlebrush was dried *in vacuo*. The polymer was characterized with $^1$H NMR. The actual charge fraction $\alpha$ was calculated by referencing P4MCL methyl protons (0.8 ppm, 30 H, marked as black open circle) compared to the imidazolium protons (9.5 ppm, marked as black open circle), i.e. $\alpha = \frac{A_{\text{imidazolium protons}}}{A_{\text{imidazolium protons}} + A_{\text{methyl protons}}/30}$, where $A$ is the integral area of proton peaks. The backbone degree of polymerization ($N_{\text{BB}}$) of the BB-Cation copolymer is 100, consistent with the 100:1 molar ratio of total monomers (ionic monomers plus macromonomers) to G3 catalyst used in the polymerization. Because G3 is a highly active ROMP catalyst, and the ring-opening metathesis polymerization is typically completed within several hundred minutes[8], this feed ratio is expected to closely determine the backbone length.

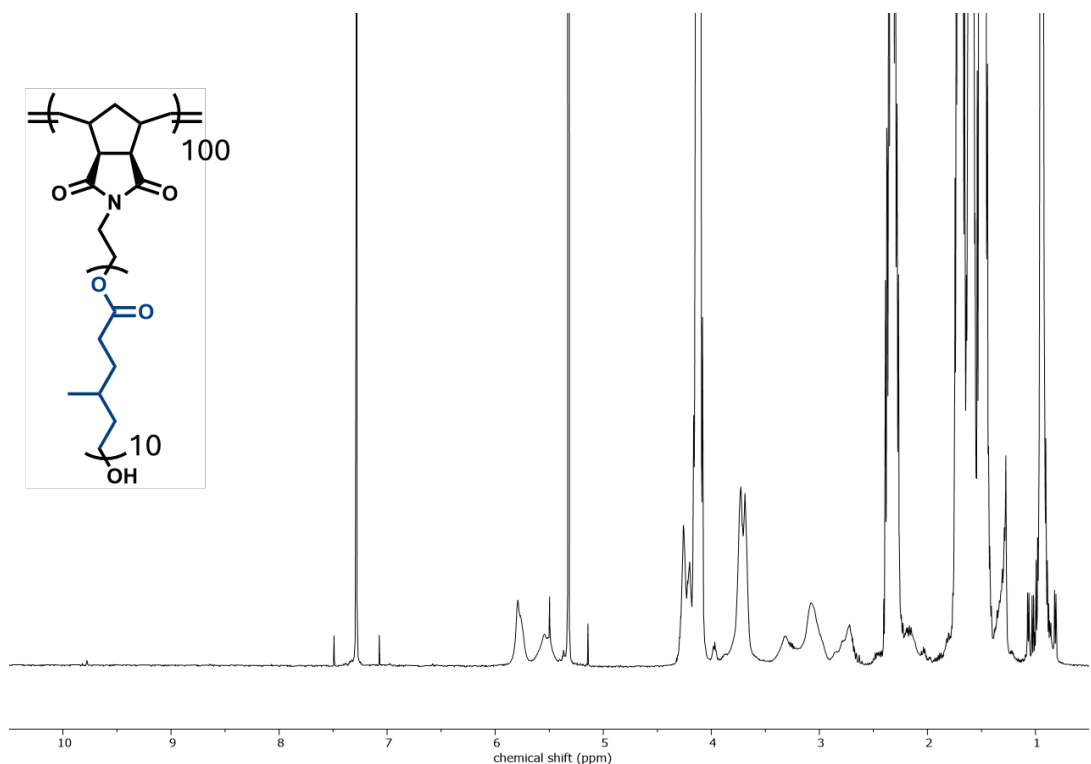

**Fig. S9.** $^1$H NMR (CDCl$_3$) of 0% charged P4MCL bottlebrush.

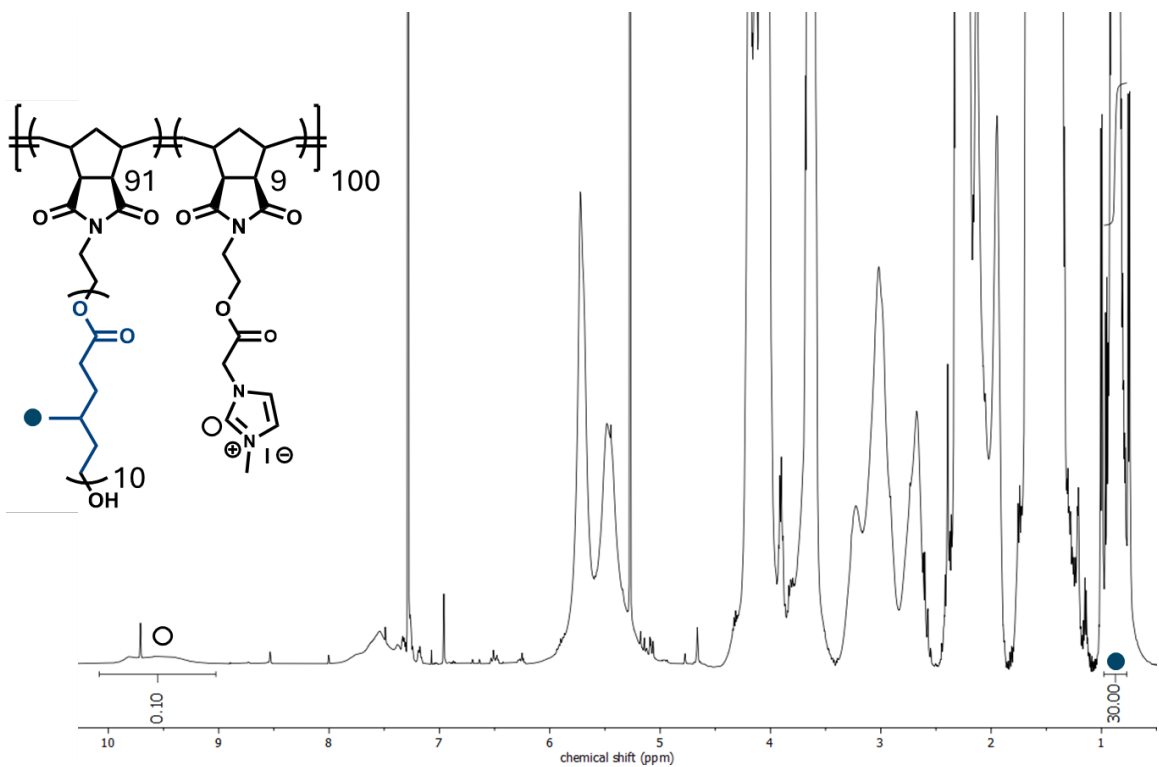

**Fig. S10.** $^1$H NMR (CDCl$_3$) of 9% charged imidazolium functionalized P4MCL bottlebrush.

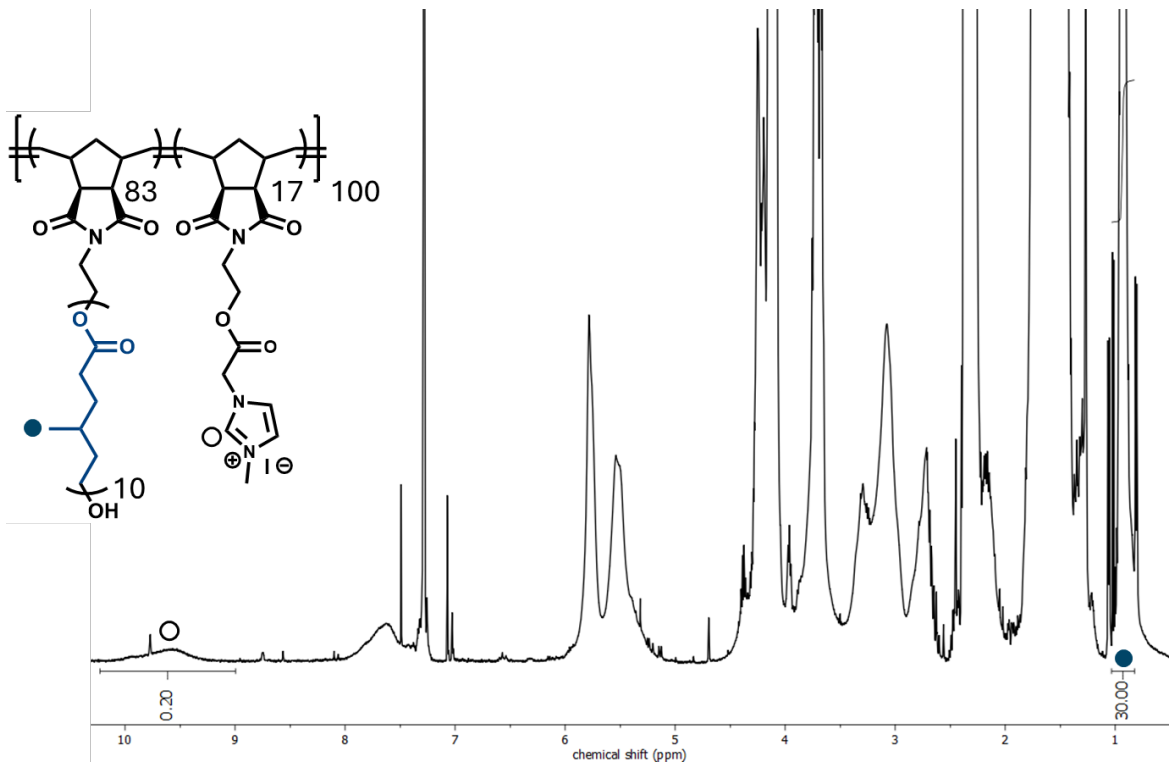

**Fig. S11.** $^1$H NMR (CDCl$_3$) of 17% charged imidazolium functionalized P4MCL bottlebrush.

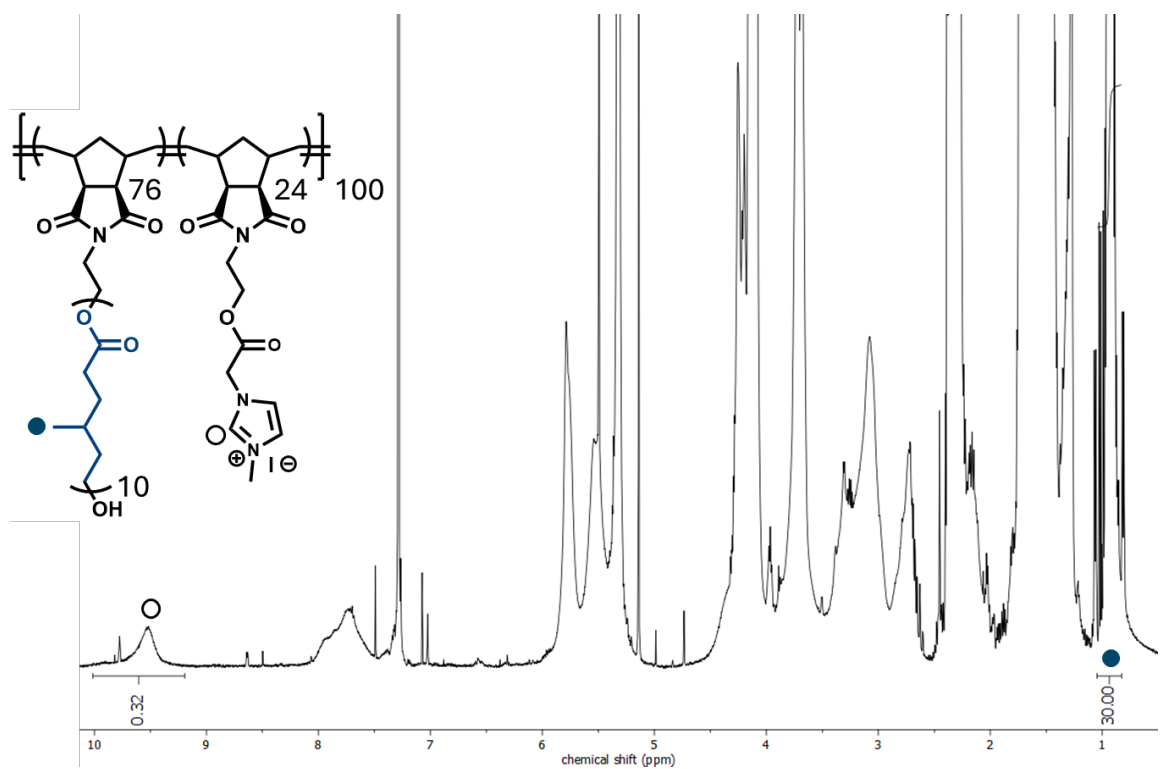

**Fig. S12.** $^1$H NMR (CDCl$_3$) of 24% charged imidazolium functionalized P4MCL bottlebrush.

*Carboxylic acid functionalized PDMS bottlebrush (BB-Anion Copolymer)*

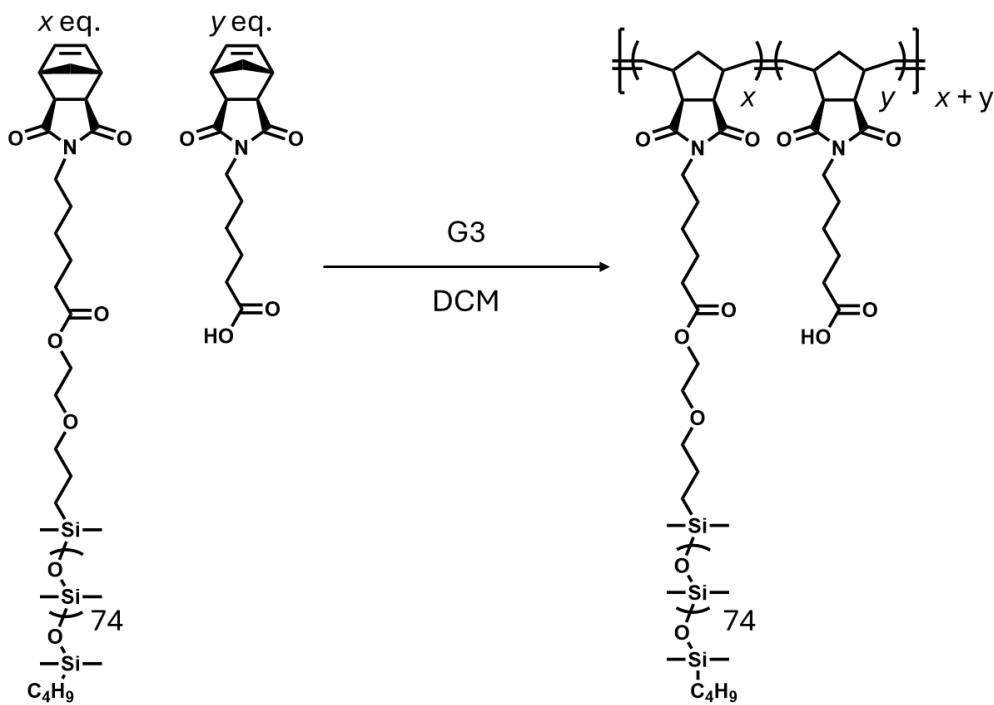

A grafting-through polymerization of PDMS macromonomer and carboxylic functionalized norbornene anhydride monomer was performed in a diluted solution of macromonomer and monomer. The monomeric charged fraction of the resulting polymer was controlled through the added ratio between PDMS macromonomer and carboxylic functionalized norbornene anhydride monomer. For example, a 40% nominal charged carboxylic functionalized bottlebrush (actual charge fraction: 43%) can be made as follows. In a nitrogen-filled glovebox, PDMS macromonomer (3 g, 0.58 mmol), and carboxylic functionalized norbornene anhydride monomer (107.1 mg, 0.39 mmol) were dissolved in anhydrous DCM. The total concentration was kept at 100 mM. In a separate container, G3 (7.0 mg, 0.010 mmol) was dissolved in 0.2 mL of anhydrous DCM before rapid injection into a vigorously stirring solution of the macromonomer and monomer. After 24 hours, the polymerization was removed from the glovebox and terminated using a large excess ethyl vinyl ether (EVE) (>100 eq) with stirring for 30 minutes. A crude $^1$H NMR of the freshly quenched reaction was taken. The disappearance of a peak at 6.3 ppm corresponding to unreacted olefin protons was observed, suggesting high reaction conversion. The resulting reaction mixture was concentrated *in vacuo* and the polymer precipitated into methanol. The carboxylic functionalized PDMS bottlebrush was dried *in vacuo*. The polymer was characterized with $^1$H NMR. The actual charge fraction was calculated by referencing methyl protons on the PDMS (0 ppm, 444 H) to the alpha methylene proton next to the carboxylic and ester (2.3 ppm), i.e. $\alpha = 1 - \frac{A_{\text{methyl protons}}/444}{A_{\text{alpha methylene proton}}/2}$. Like BB-Cation copolymer, the backbone degree of polymerization ($N_{BB}$) of the BB-Anion copolymer is also 100.

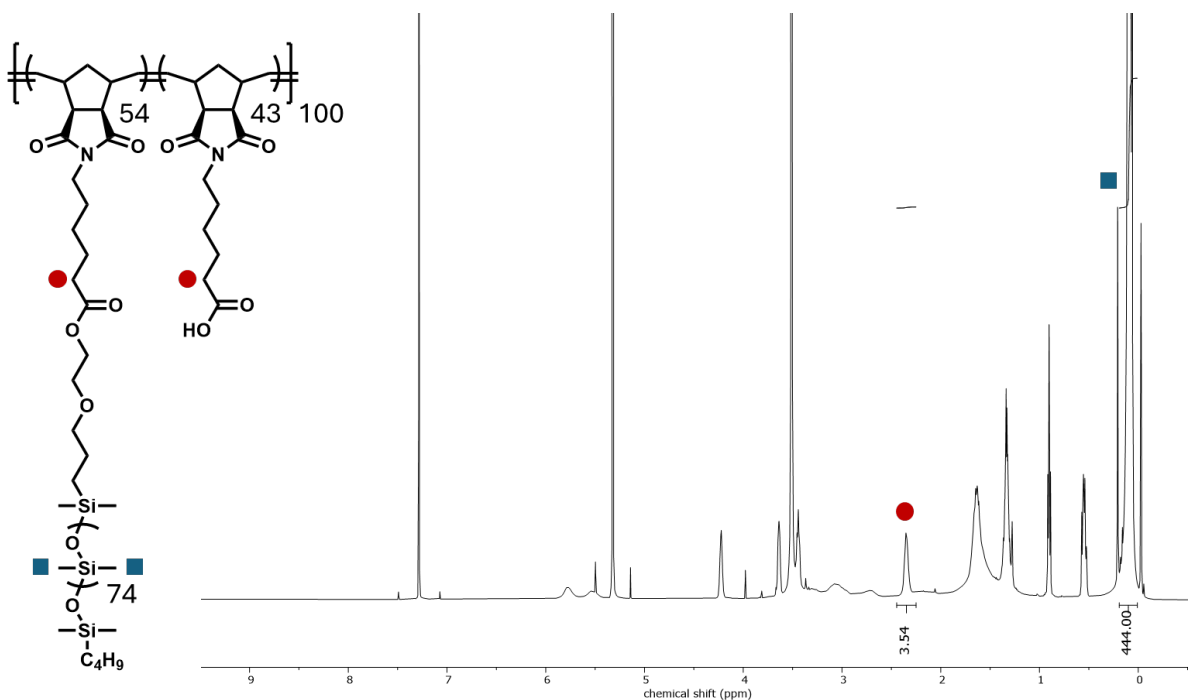

**Fig. S13** $^1$H NMR (CDCl$_3$) of 43% carboxylic functionalized PDMS bottlebrush.

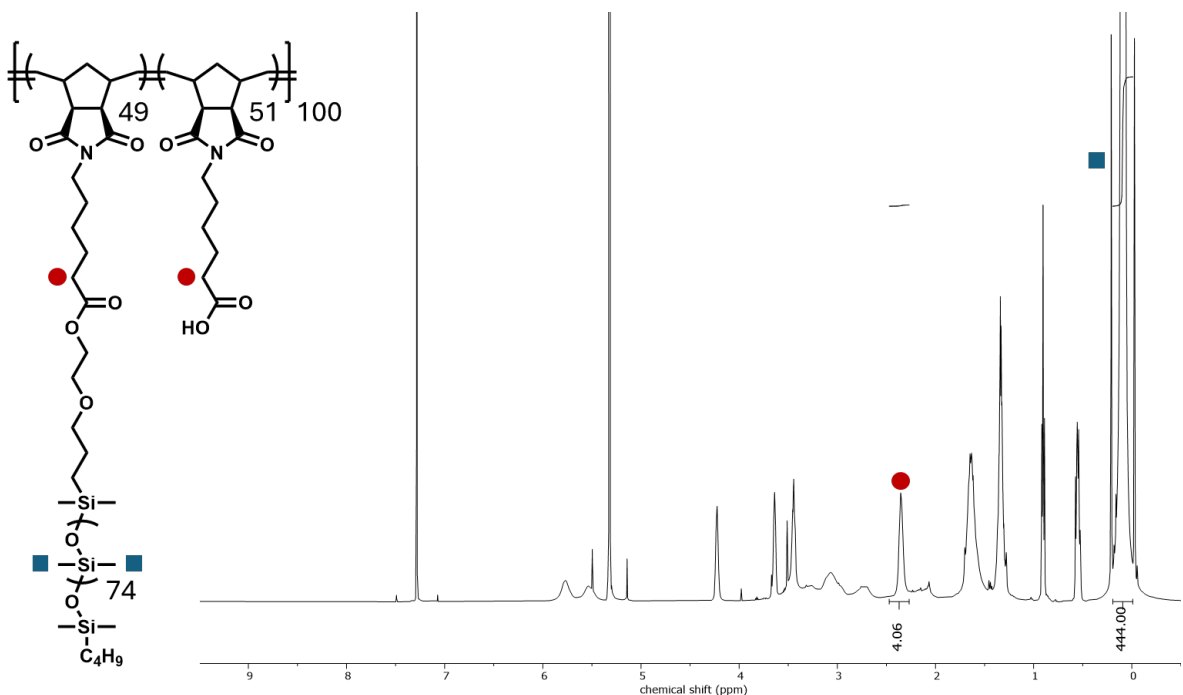

**Fig. S14.** $^1$H NMR (CDCl$_3$) of 51% carboxylic functionalized PDMS bottlebrush.

| BB-Cation | | BB-Anion | |
|---|---|---|---|
| Nominal Charge Fraction (%) | Actual Charge Fraction $\alpha$ (%) | Nominal Charge Fraction (%) | Actual Charge Fraction $\alpha$ (%) |
| 0 | 0 | 40 | 43 |
| 10 | 9 | 50 | 51 |
| 20 | 17 | | |
| 30 | 24 | | |

**Table S1** Summary of charge fraction for as-synthesized BB-Cation and BB-Anion

**Cross-linking.** The as-synthesized BB-Cation and BB-Anion polymers were mixed with cross-linkers and fabricated on substrates to cross-link under ultra-violet (UV) light. For example, a 10% nominal charged imidazolium functionalized bottlebrush (actual charge fraction: 9%) can be cross-linked as follows. In a clean vial, BB-Cation (10% nominal charge fraction, $\alpha$ = 9%, 1 g, 0.007 mmol), and cross-linkers (60.27 mg, 0.105 mmol) were dissolved in ~ 15 mL DCM. After mixing, the DCM evaporated and the mixture was dried thoroughly in a vacuum oven for 48 h. The as dried mixture was then bladed-coated on substrates (aluminum plate for rheology testing and porous carbon electrode for adhesion testing, Supporting Video 1 and 2) to cross-link under UV light (365 nm). The molecular structures of the cross-linking agents are shown in Fig. S15.

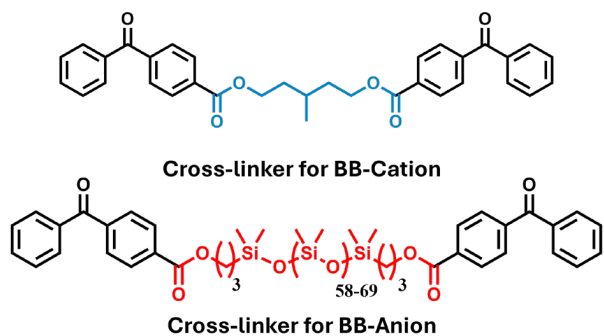

**Fig. S15.** Chemical structures of cross-linker for BB-Cation and BB-Anion bottlebrush polymers. The number of repeat units of the cross-linker for BB-Anion bottlebrush polymers may slightly vary.

**Methods**

**Nuclear Magnetic Resonance (NMR).** All $^1$H, $^{13}$C NMR, and 2D-NMR spectra were collected using a Bruker Avance NEO 500 MHz equipped with a CryoProbe Prodigy BBO probe with z-axis pulsed field gradient (PFG).

**Size-Exclusion Chromatography (SEC).** SEC was performed on a Waters Alliance High-performance liquid chromatography (HPLC) System equipped with a 2690 Separation Module, or a Waters Alliance HPLC system with an Agilent PLgel 5 μm MiniMIX-D column. The former uses chloroform with 0.25% triethylamine, while the latter uses THF as the eluent. For the former, the refractive index from a Waters 2410 Differential Refractometer detector was used to estimate the molar mass and dispersity relative to linear polystyrene standards. For the latter, refractive index traces from a Waters 2414 detector were used for molecular weight determination using polystyrene calibration standards (Agilent Technologies).

## 2. Differential scanning calorimetry of BB-Anion and heat capacity of BB-Cation

Figure S16 plots heat-flow traces obtained by Differential Scanning Calorimetry (DSC) for BB-Anion at various charge fractions. In all cases, the $T_g$ values are not shown because they are below −100 °C[9].

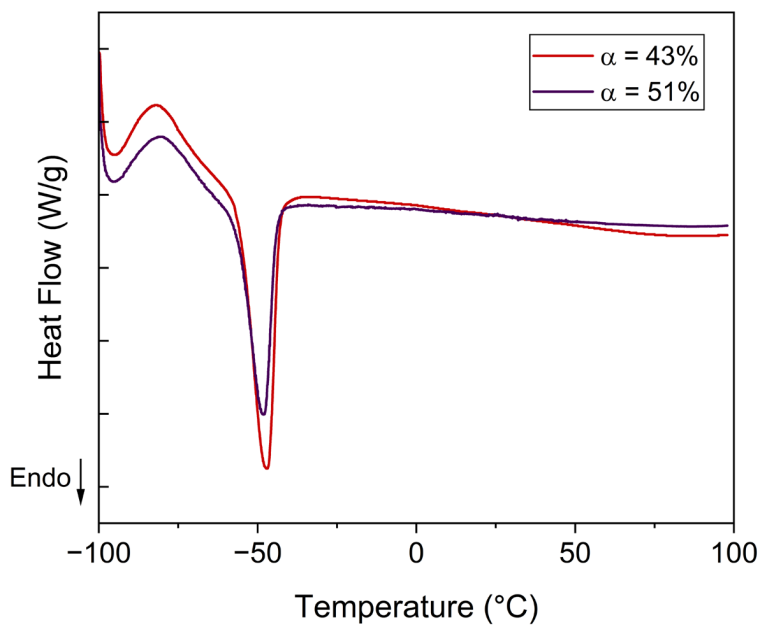

**Fig. S16.** DSC results of BB-Anion with different charge fractions.

## 3. Supplementary mechanical engineering characterizations for BB-Cation and BB-Anion

Figure S17 shows tan $\delta = G''/ G'$ of cross-linked BB-Cation elastomer with different charge fractions (See Methods for details). Figure S18 a and b show shear modulus and tan $\delta$ calculated from the measured moduli of cross-linked BB-Anion. Notably, in Fig. S18a the BB-Anion exhibits an apparent relaxation-like feature in the high reduced-frequency region, manifested as a decrease in $G''$ with increasing frequency. We attribute this behavior to minor solvent loss during the high-temperature portion of the time–temperature superposition (TTS) protocol. For TTS, isothermal frequency sweeps from 100 to 0.1 rad s$^{-1}$ at a constant strain of 1% were obtained at multiple temperatures (−10 °C to 100 °C), and master curves were constructed using the Williams–Landel–Ferry (WLF) relation. BB-Anion films were converted from terminal COOH to the potassium carboxylate form by doping in a KOH solution prepared in THF/DI water (1:1), which can leave residual THF trapped in the network. During rheological frequency sweeps collected from −10 °C to 100 °C and shifted to a room-temperature (25 °C) master curve, the highest-temperature measurements which map to the highest reduced-frequency region at reference temperature (25 °C) may promote de-swelling/evaporation of residual THF. This de-swelling reduces viscous dissipation, leading to a lower measured $G''$ and hence the observed downturn in the high reduced-frequency regime. Note that this shift method also caused oscillation of tan $\delta$ data especially in the low frequency region ($\alpha$ = 17% in Fig. S17 and $\alpha$ = 43% in Fig. S18b), which can be attributed to an error in the temperature shift parameter. Figure S19 shows the linear region of strain-stress curves of BB-Cation with different charge fractions. Figure S20 shows the linear region of Fig. 3d.

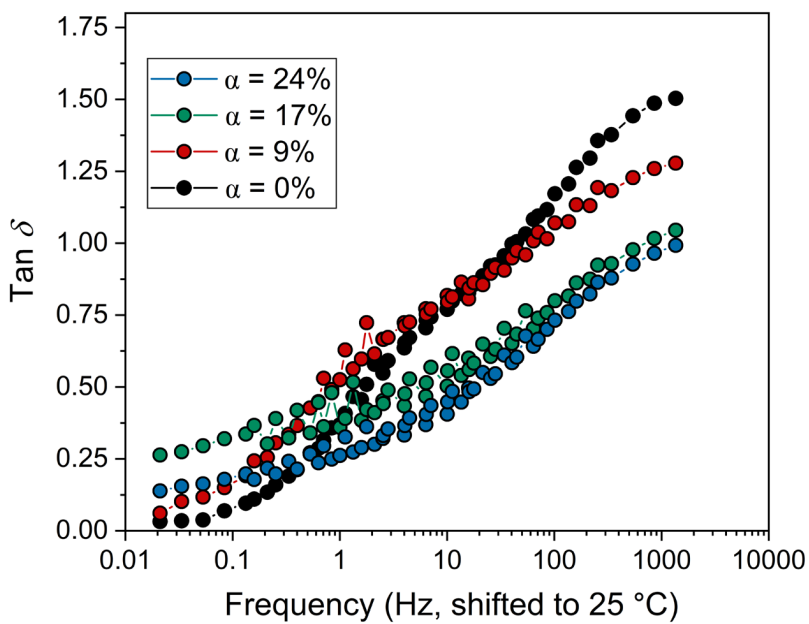

**Fig. S17.** Tan $\delta$ of BB-Cation with different charge fractions.

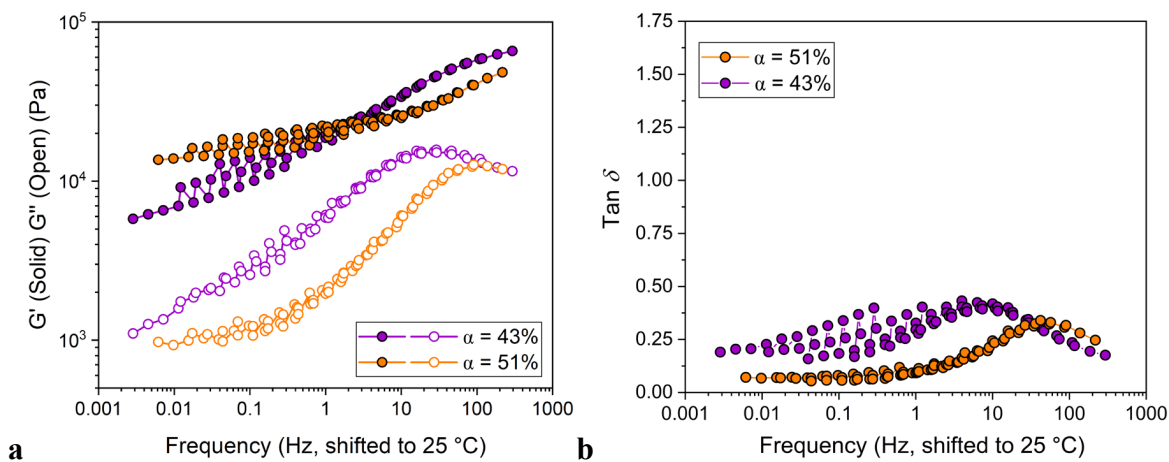

**Fig. S18.** Rheological data (a) and tan $\delta$ (b) of cross-linked BB-Anion with different charge fractions. In a, the closed and open symbols represent $G'$ (storage) and $G''$ (loss) components respectively.

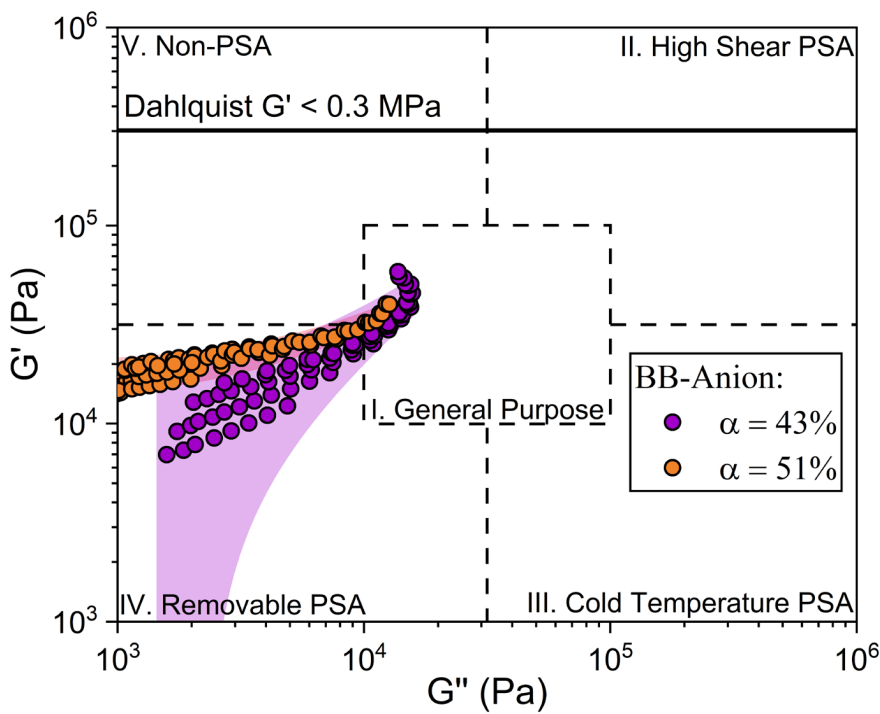

**Fig. S19.** Viscoelastic windows for BB-Cation and BB-Anion with different charge fractions. The storage ($G'$) and loss ($G''$) moduli over the range of $\omega \approx 0.01$–$100$ rad s$^{-1}$ are plotted, with the shaded background indicating the 95% confidence interval. The Dahlquist criterion for efficient contact ($G' < 0.3$ MPa) is also shown.

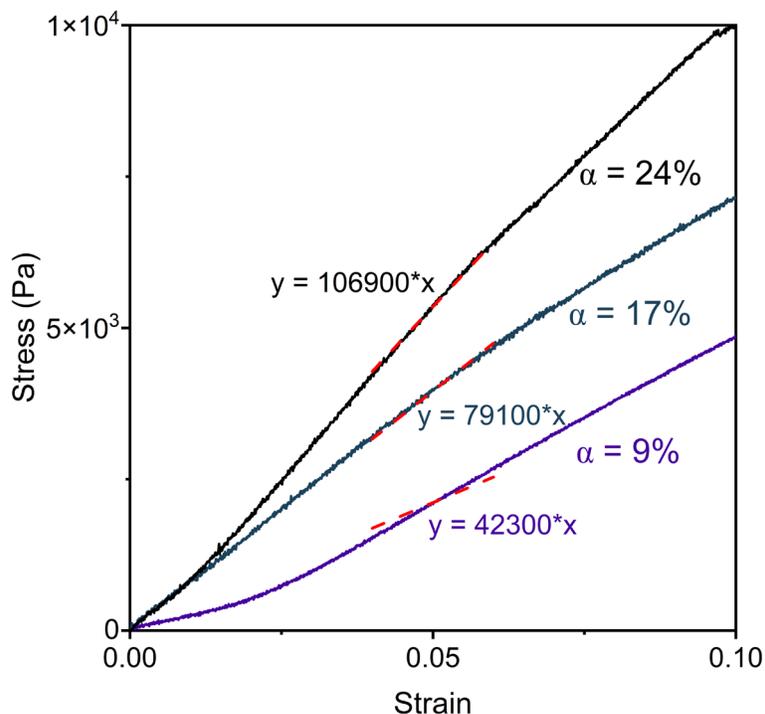

**Fig. S20.** Linear region of strain-stress curves of BB-Cation with different charge fractions. The measured Young's moduli of BB-Cation with 9%, 17%, and 24% are 42, 79, and 107 kPa, respectively. The red dashed lines and corresponding equations represent the linear fits to the stress–strain curves. The fitting was performed over the strain range of 0.04–0.06, and a zero-intercept constraint was imposed on all three fitted curves. Notably, the 9% BB-Cation trace shows apparent low-strain stiffening, likely due to the very soft and tacky nature of the elastomer. After mounting in the fixture, the film is difficult to keep fully suspended and free of slack; at the onset of tensile loading, the sample primarily undergoes straightening, which artificially reduces the initial slope. As a result, the small initial slope, together with the enforced zero-intercept condition, causes the fitted curve for 9% BB-Cation to deviate from the experimental data.

## 4. EIS of BB-Cation

Figures S21-24 summarize the EIS measurement results and equivalent circuit fitting curves of BB-Cation with different charge fractions. For each charge fraction case, 2 repeated measurements were performed on 2 different samples. Table S2 lists the EIS equivalent circuit parameters fitting results. The fitting results show that $R_1$, which represents the contact resistance at the electrode/polymer interface, increases with charge fraction. This increase indicates poorer interfacial contact, likely arising from the higher viscosity of the polymer precursor and the associated fabrication challenges. As described in the main text, $R_2$ is related to ion conductivity of the polymer. The ion conductivity $\sigma$ of the ion-containing elastomers can be extracted from $\sigma =$
20

$\frac{t}{R_2 A}$, where $t$ is the elastomer thickness and $A$ is the contact area with the ITO-coated electrode. The high-frequency polarization of the ion-containing network, represented by $Q_1$, decreases with increasing charge fraction. This trend is likely due to the higher effective cross-link density, which constrains chain relaxation. Generally, $Q_2$, which is related to the electric double layer at the electrode/elastomer interface, increases with charge fraction (due to increased ion concentration). However, the driving voltage for our EIS measurement is 100 mV, which might be too small to drive all the ions the to the electrode/polymer interface, leading to a decrease in $Q_2$ when the charge fraction is 17% and 24%. The ion conductivity of BB-Cation with different charge fractions is shown in Fig. S25.

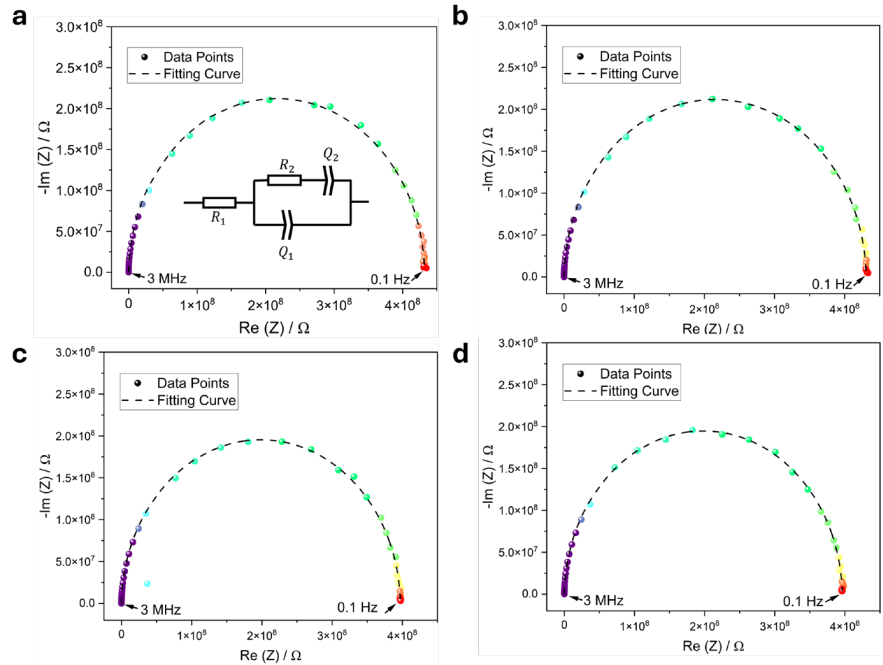

**Fig. S21.** Electro-chemical characterization of 0% charge fraction BB-Cation. **a,b,c** Nyquist impedance plots of 0% charge fraction BB-Cation, in which (**a**) and (**b**) were collected on a same sample, and (**c**) and (**d**) were collected on another sample. The color of the data points indicates the measurement frequency of the circuit, ranging from purple (3 MHz) to red (0.1 Hz).

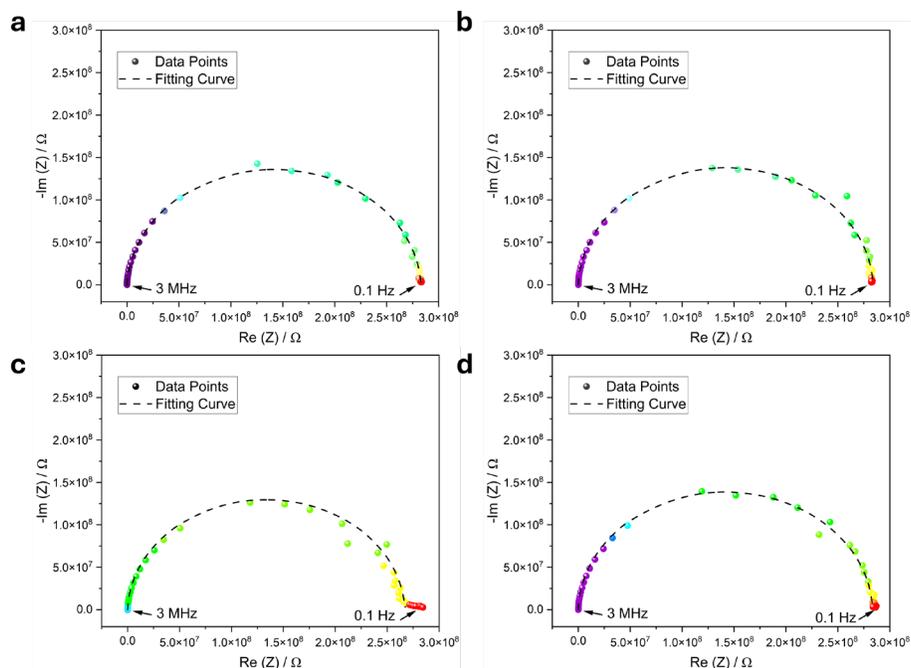

**Fig. S22.** Electro-chemical characterization of 9% charge fraction BB-Cation. **a,b,c,d** Nyquist impedance plots of 9% charge fraction BB-Cation, in which (**a**) and (**b**) were collected on a same sample, and (**c**) and (**d**) were collected on another sample. The color of the data points indicates the measurement frequency of the circuit, ranging from purple (3 MHz) to red (0.1 Hz).

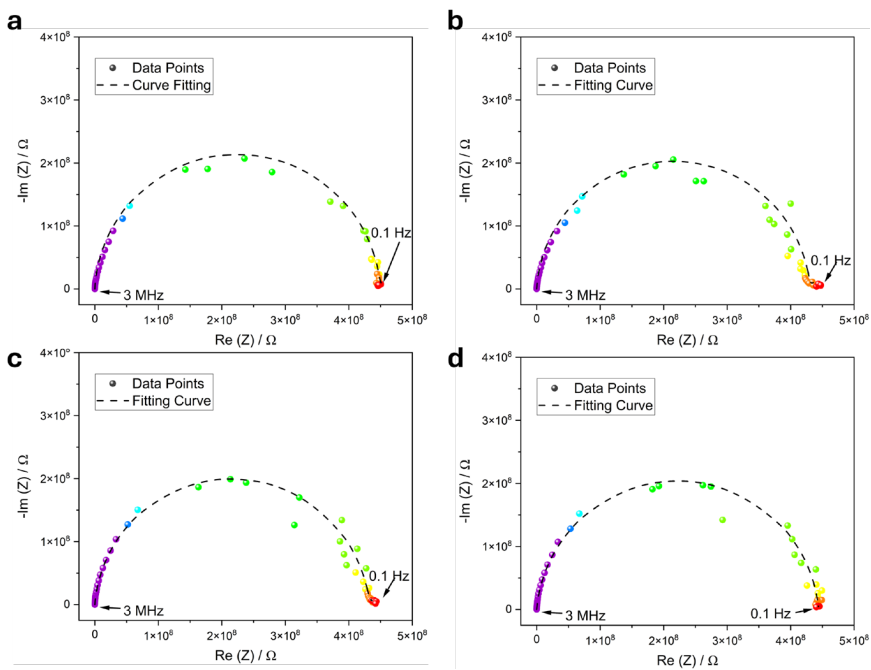

**Fig. S23.** Electro-chemical characterization of 17% charge fraction BB-Cation. **a,b,c,d** Nyquist impedance plots of 17% charge fraction BB-Cation, in which (**a**) and (**b**) were collected on a same sample, and (**c**) and (**d**) were collected on another sample. The color of the data points indicates the measurement frequency of the circuit, ranging from purple (3 MHz) to red (0.1 Hz).

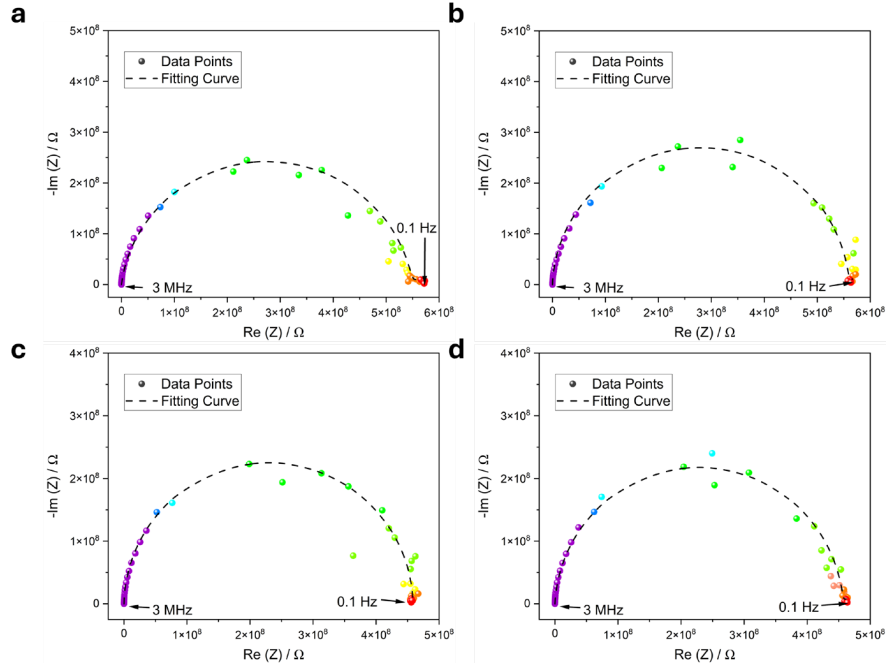

**Fig. S24.** Electro-chemical characterization of 24% charge fraction BB-Cation. **a,b,c,d** Nyquist impedance plots of 24% charge fraction BB-Cation, in which (**a**) and (**b**) were collected on a same sample, and (**c**) and (**d**) were collected on another sample. The color of the data points indicates the measurement frequency of the circuit, ranging from purple (3 MHz) to red (0.1 Hz).

|  | **0% BB-Cation** | **9% BB-Cation** | **17% BB-Cation** | **24% BB-Cation** |
|---|---|---|---|---|
| $R_1$/ohm | 107.9 ± 15.9 | 199.8 ± 13.3 | 227.4 ± 15.4 | 321.1 ± 16.2 |
| $R_2$/ohm | (4.13 ± 0.19) × 10$^8$ | (2.79 ± 0.07) × 10$^8$ | (4.37 ± 0.09) × 10$^8$ | (5.05 ± 0.48) × 10$^8$ |
| $Q_1$/(F*s$^{\alpha_1-1}$) | (1.90 ± 0.08) × 10$^{-11}$ | (1.82 ± 0.04) × 10$^{-11}$ | (1.56 ± 0.05) × 10$^{-11}$ | (1.13 ± 0.13) × 10$^{-11}$ |
| $\alpha_1$ | 0.99 ± 0.01 | 0.98 ± 0.01 | 0.96 ± 0.01 | 0.96 ± 0.03 |
| $Q_2$/(F*s$^{\alpha_2-1}$) | (4.25 ± 1.14) × 10$^{-7}$ | (5.22 ± 0.41) × 10$^{-7}$ | (2.07 ± 1.42) × 10$^{-7}$ | (1.52 ± 0.64) × 10$^{-7}$ |
| $\alpha_2$ | 1.00 ± 0.01 | 0.89 ± 0.07 | 0.57 ± 0.27 | 0.61 ± 0.23 |
| Film Thickness/mm | 0.28 ± 0.01 | 0.37 ± 0.01 | 0.44 ± 0.15 | 0.49 ± 0.05 |
| Ion conductivity /(S*cm$^{-1}$) | (8.12 ± 0.22) × 10$^{-10}$ | (1.68 ± 0.05) × 10$^{-9}$ | (1.26 ± 0.05) × 10$^{-9}$ | (1.25 ± 0.25) × 10$^{-9}$ |

**Table S2.** EIS equivalent circuit parameters fitting results of BB-Cation with different charge fractions. The equivalent circuit model is shown in Fig. S21a inset.

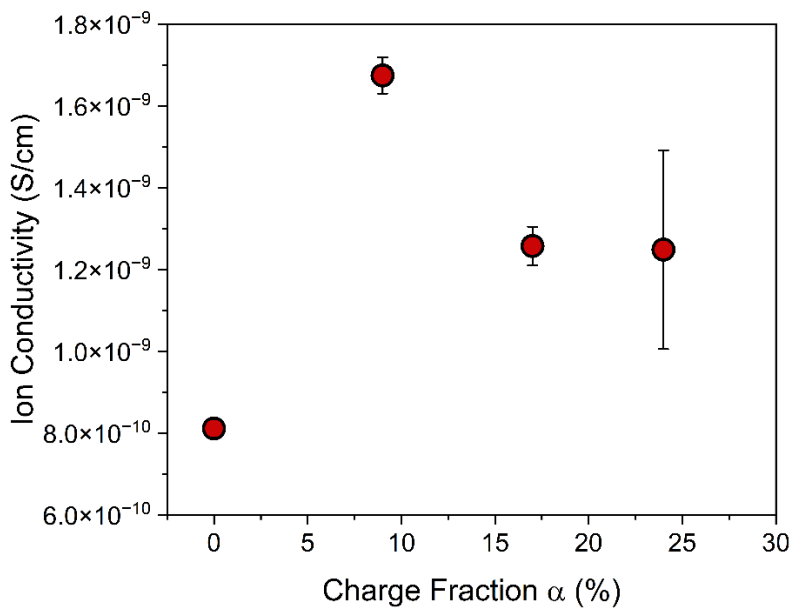

**Fig. S25.** Ion conductivity of BB-Cation with different charge fractions.

## 5. EIS of BB-Anion

Figures S26-27 summarize the EIS measurement results and equivalent circuit fitting curves of BB-Anion with different charge fractions. For each charge fraction case, 2 repeated measurements were performed on 2 different samples. Table S3 lists the EIS equivalent circuit parameters fitting results. The high-frequency polarization of the ion-containing network, represented by $Q_1$, decreases with increasing charge fraction. This trend is likely due to the higher effective cross-link density, which constrains chain relaxation. In contrast to BB-Cation, the ion conductivity extracted from $R_2$ of BB-Anion increases with charge fraction. Because of the long PDMS chains (Fig. S13 and 14), steric hindrance may impede ionic interactions between fixed ions on the polymer backbone and mobile ions in the matrix. As a result, fewer mobile ions may be trapped by fixed charges during migration, leaving a larger fraction of charge carriers available for transport and thus increasing ionic conductivity. Moreover, the increasing trend of $Q_2$ (the electric double layer at the electrode/elastomer interface) with charge fraction, is observed in Table S3.

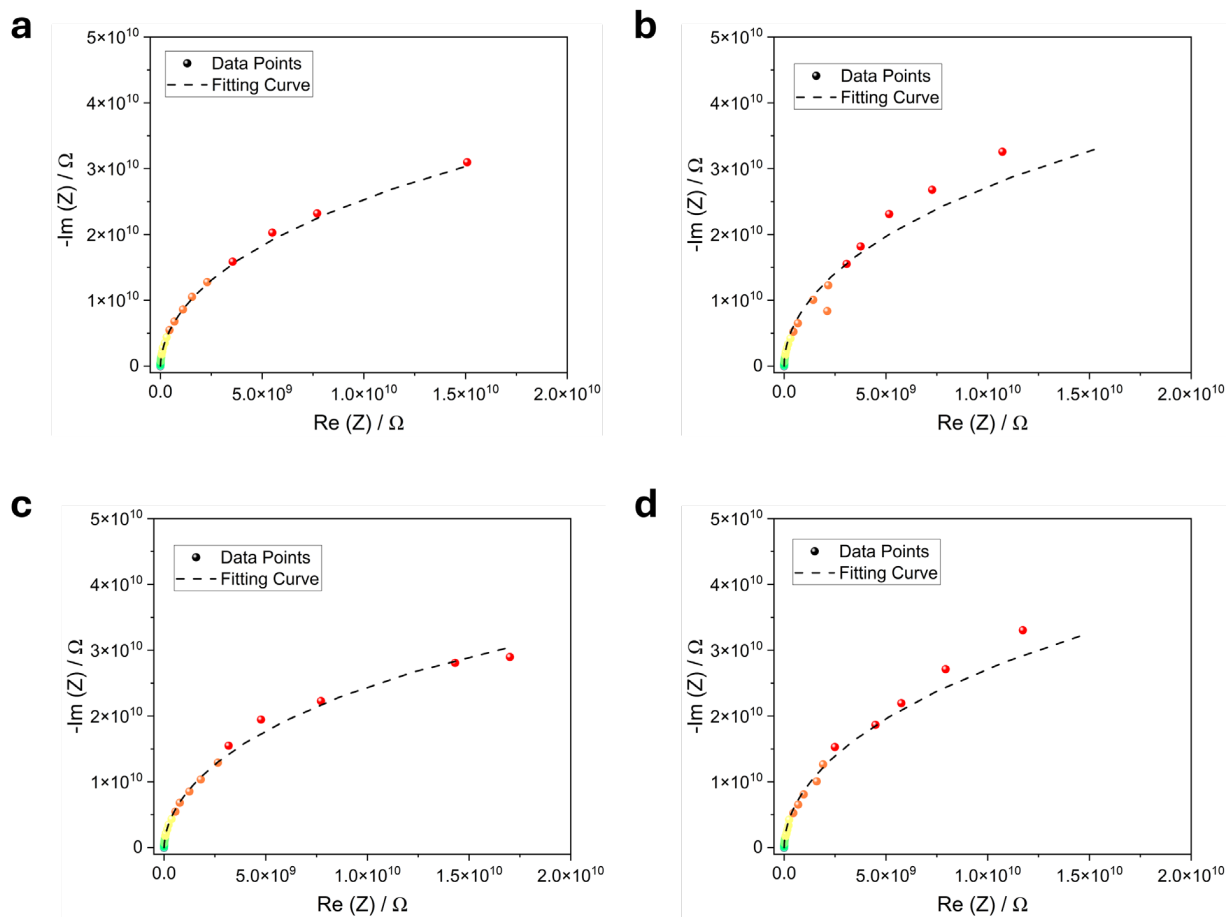

**Fig. S26.** Electro-chemical characterization of 43% charge fraction BB-Anion. **a,b,c,d** Nyquist impedance plots of 43% charge fraction BB-Cation, in which (**a**) and (**b**) were collected on a same sample, and (**c**) and (**d**) were collected on another sample. The color of the data points indicates the measurement frequency of the circuit, ranging from purple (3 MHz) to red (0.1 Hz).

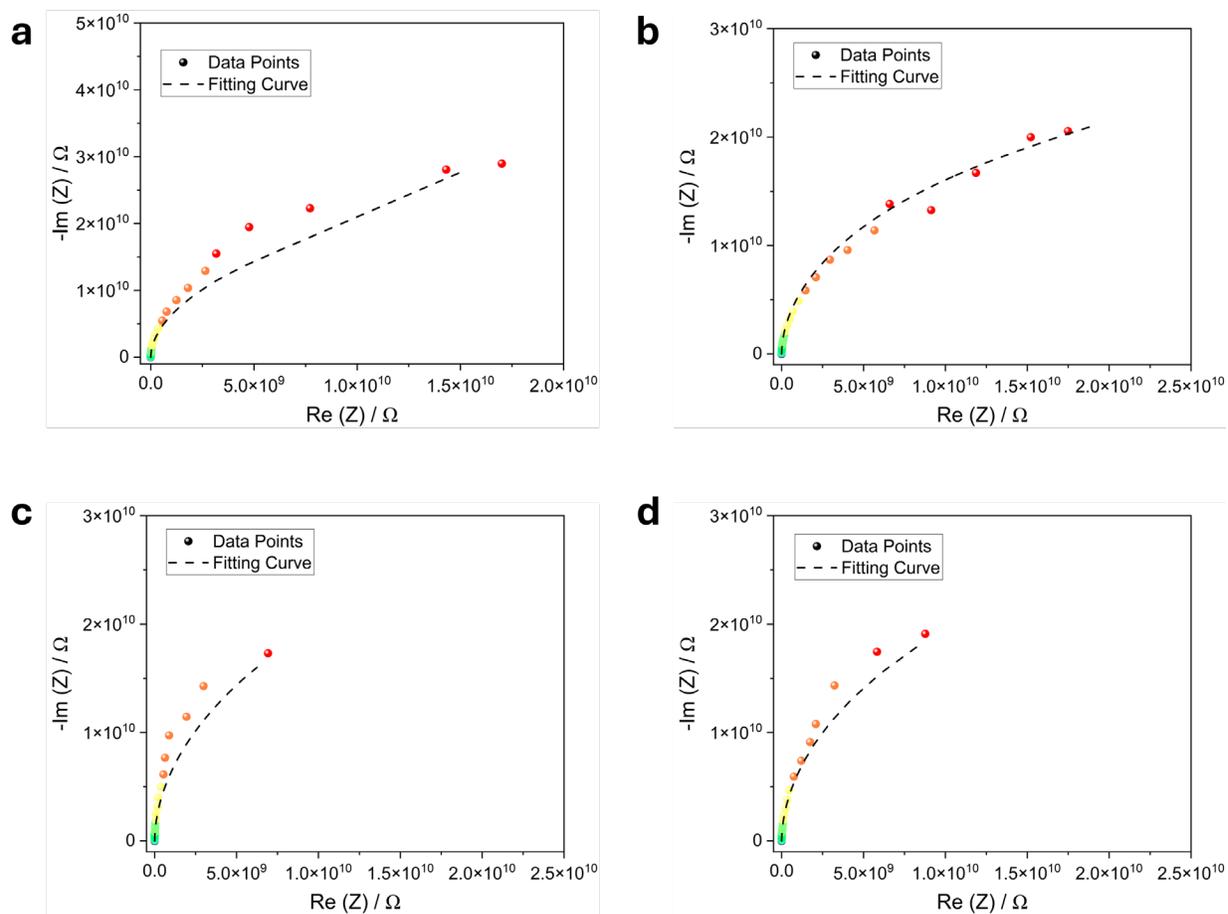

**Fig. S27.** Electro-chemical characterization of 51% charge fraction BB-Anion. **a,b,c,d** Nyquist impedance plots of 51% charge fraction BB-Cation, in which (**a**) and (**b**) were collected on a same sample, and (**c**) and (**d**) were collected on another sample. The color of the data points indicates the measurement frequency of the circuit, ranging from purple (3 MHz) to red (0.1 Hz).

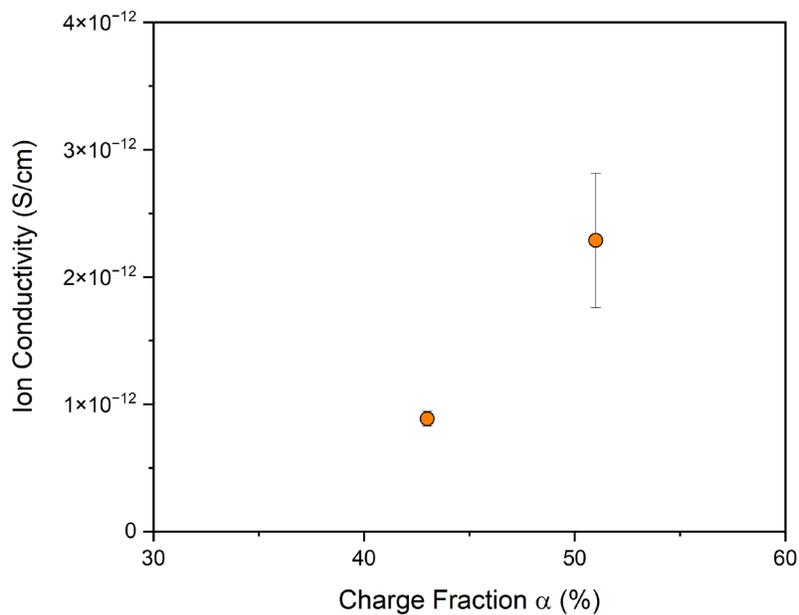

**Fig. S28.** Ion conductivity of BB-Anion with different charge fractions.

|  | **43% BB-Anion** | **51% BB-Anion** |
|---|---|---|
| $R_1$/ohm | 66.2 ± 6.6 | 63.0 ± 7.8 |
| $R_2$/ohm | $(7.64 \pm 0.49) \times 10^{10}$ | $(3.11 \pm 0.71) \times 10^{10}$ |
| $Q_1/(F*s^{\alpha_1 -1})$ | $(3.67 \pm 0.05) \times 10^{-11}$ | $(3.09 \pm 0.19) \times 10^{-11}$ |
| $\alpha_1$ | 1.00 ± 0.01 | 0.99 ± 0.01 |
| $Q_2/(F*s^{\alpha_2 -1})$ | $(1.07 \pm 0.26) \times 10^{-10}$ | $(6.79 \pm 0.30) \times 10^{-7}$ |
| $\alpha_2$ | 0.92 ± 0.06 | 0.57 ± 0.05 |
| Film Thickness/mm | 0.12 ± 0.01 | 0.12 ± 0.01 |
| Ion conductivity /(S*cm$^{-1}$) | $(8.86 \pm 0.58) \times 10^{-13}$ | $(2.29 \pm 0.53) \times 10^{-12}$ |

**Table. S3.** EIS equivalent circuit parameters fitting results of BB-Anion with different charge fractions. The equivalent circuit model is shown in Fig. 3a and b inset.

## 6. EIS equivalent circuit parameters fitting results of ion-containing bottlebrush elastomer heterojunction

| Voltage/V | 0 | 0.5 | 1 | 2 |
|---|---|---|---|---|
| $R_1$/ohm | $2.04 \times 10^2$ | $1.47 \times 10^2$ | $2.47 \times 10^2$ | $1.11 \times 10^2$ |
| $Q_1/(F*s^{\alpha_1 -1})$ | $1.81 \times 10^{-11}$ | $1.79 \times 10^{-11}$ | $1.75 \times 10^{-11}$ | $1.77 \times 10^{-11}$ |
| $\alpha_1$ | 0.97 | 0.97 | 0.98 | 0.98 |
| $R_2$/ohm | $9.90 \times 10^9$ | $9.69 \times 10^9$ | $9.79 \times 10^9$ | $9.79 \times 10^9$ |
| $Q_2/(F*s^{\alpha_2 -1})$ | $5.55 \times 10^{-10}$ | $6.60 \times 10^{-11}$ | $3.61 \times 10^{-11}$ | $2.70 \times 10^{-11}$ |
| $\alpha_2$ | 0.33 | 0.21 | 0.26 | 0.25 |
| $Q_3/(F*s^{\alpha_3 -1})$ | $5.56 \times 10^{-12}$ | $6.80 \times 10^{-12}$ | $7.95 \times 10^{-12}$ | $8.14 \times 10^{-12}$ |
| $\alpha_2$ | 1.00 | 1.00 | 1.00 | 1.00 |
| $R_3$/ohm | $1.47 \times 10^4$ | $2.54 \times 10^6$ | $3.58 \times 10^6$ | $5.92 \times 10^6$ |

**Table S4.** EIS equivalent circuit parameters fitting results of ion-containing bottlebrush elastomer heterojunction formed in between 17% BB-Cation and 51% BB-Anion. The equivalent circuit model is shown in Fig. 4c inset.

## 7. Adhesion measurement

Figure S29 shows a representative sample for adhesion measurement and a sample fabricated on a white paper. The film is roughly 6 cm long, 1 cm wide and 500 μm thick. Figure S30 shows an example of raw data (17% BB-Cation with Linear-Anion) collected in adhesion tests. The adhesion test has three stages: 1) approach stage: the upper fixture initially separated from bottom fixture by 1 mm was then lowered at 0.01 mm s$^{-1}$ until contact was established; 2) dwelling stage: the samples were held in contact for 300 s under a constant normal load of 200 mN regulated by a proportional–integral–derivative (PID) controller to allow interfacial equilibration; 3) separation stage: the upper fixture was raised at the same rate (0.01 mm s$^{-1}$) back to its original position. The force versus time data and displacement versus time data were then replotted as a JKR curve for adhesion analysis. The design of the fixture is shown in Fig. S31. Figure S32 shows JKR curves collected from adhesion measurement of BB-Cation versus Linear-Anion under different voltages.

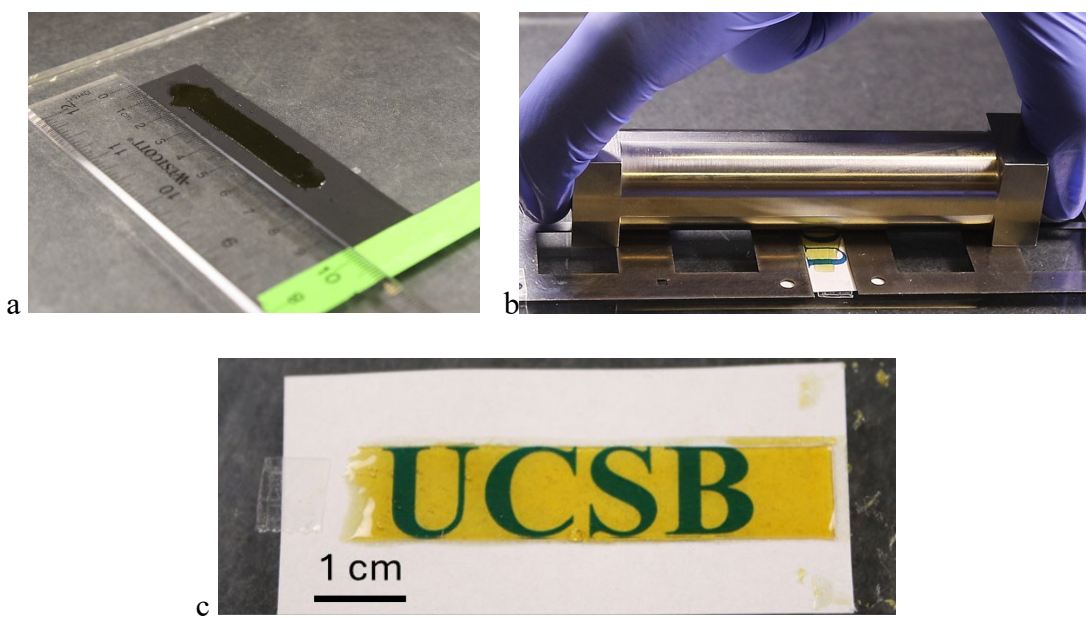

**Fig. S29.** a, A photograph of a cross-linked BB-Cation (charge fraction: 10%) film on a porous carbon electrode. b, A photograph of blade-coating fabrication of BB-Cation on a white paper. c, Photograph of a cross-linked BB-Cation elastomer film on white paper, demonstrating the smooth, defect-free surface morphology of the bottlebrush elastomer.

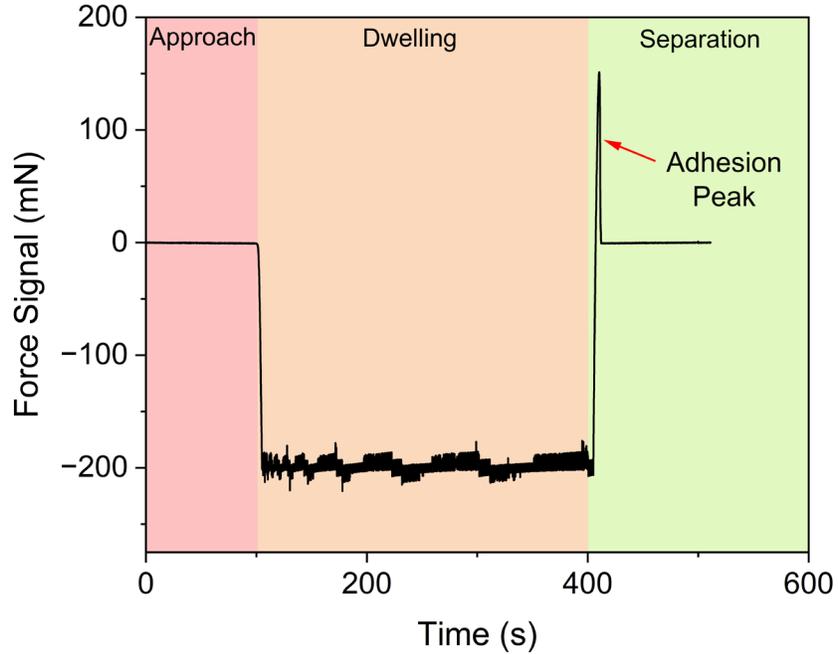

**Fig. S30.** An example of raw force versus time data collected in a representative adhesion test. There are three distinct stages. Approach stage: the upper fixture initially separated with bottom fixture for 1 mm was then lowered at 0.01 mm s$^{-1}$ until contact was established; Dwelling: The samples were held in contact for 300 s under a constant normal load of 200 mN regulated by a proportional–integral–derivative (PID) controller to allow interfacial equilibration; Separation: the upper fixture was raised at the same rate (0.01 mm s$^{-1}$) back to its original position.

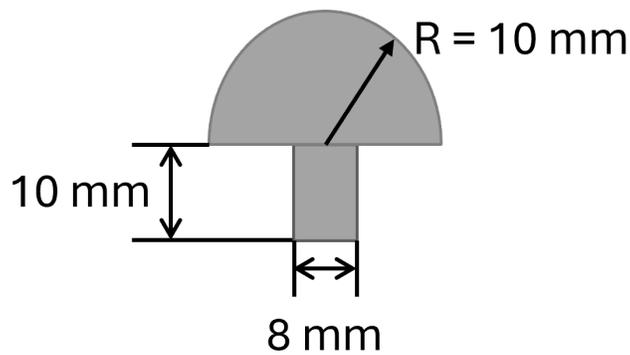

**Fig. S31.** 2D schematic showing the geometry of the custom-designed fixture. The thickness of the fixture in the third dimension is 10 mm.

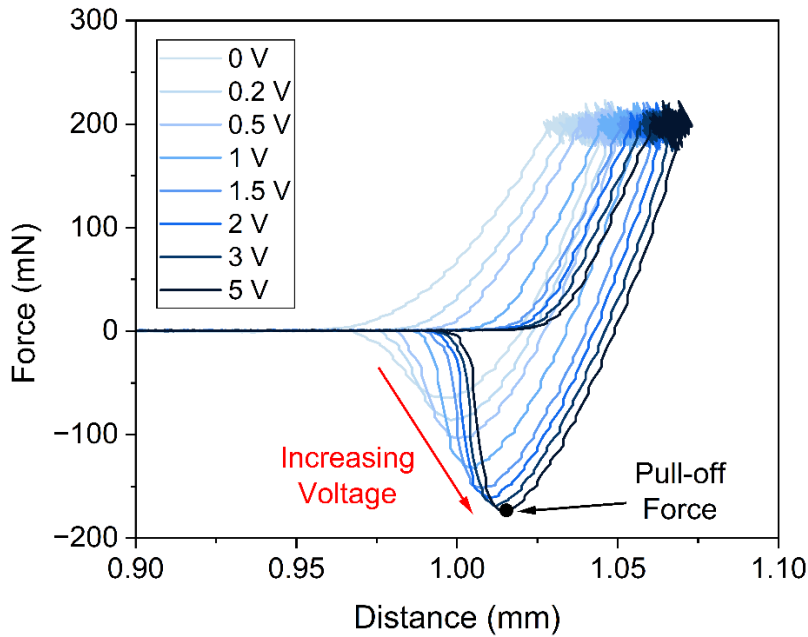

**Fig. S32.** JKR curves collected from adhesion measurement of BB-Cation ($\alpha$ = 17%) versus Linear-Anion under different voltages. The pull-off force (voltage = 5 V) is marked in the figure (black dot) and indicated by the black arrow.

## 8. Intrinsic adhesions and on/off ratio of different electroadhesive pairs

The intrinsic adhesion of BB-Cation with different charge fractions against poly[1-(2-acryloyloxyethyl)-3-butylimidazolium] bis(trifluoromethane)sulfonimide (Linear-Anion), 43% charge fraction BB-Anion, and 51% charge fraction BB-Anion is plotted in Fig. S33. The experiment setup has been described in Methods section and shown in Fig. 5c. The geometry of the fixture is shown in Fig. S31. We used pull-off force to calculate the work of adhesion: $F_{\text{pull-off}} = \frac{3}{2}\pi R W_{\text{adh}}$, where $F_{\text{pull-off}}$ is the measured pull-off force (the force at which the two fixtures are just separated, indicated by the black arrow in Fig. S32), $R$ is the radius of the fixture and $W_{\text{adh}}$ is the work of adhesion. In general, the intrinsic adhesion decreases with increasing charge fraction. Notably, the unannealed 24% charge-fraction BB-Cation exhibits relatively high intrinsic adhesion, likely due to its rough surface (Fig. S36a), which may generate more asperity contacts and thus a larger effective contact area.

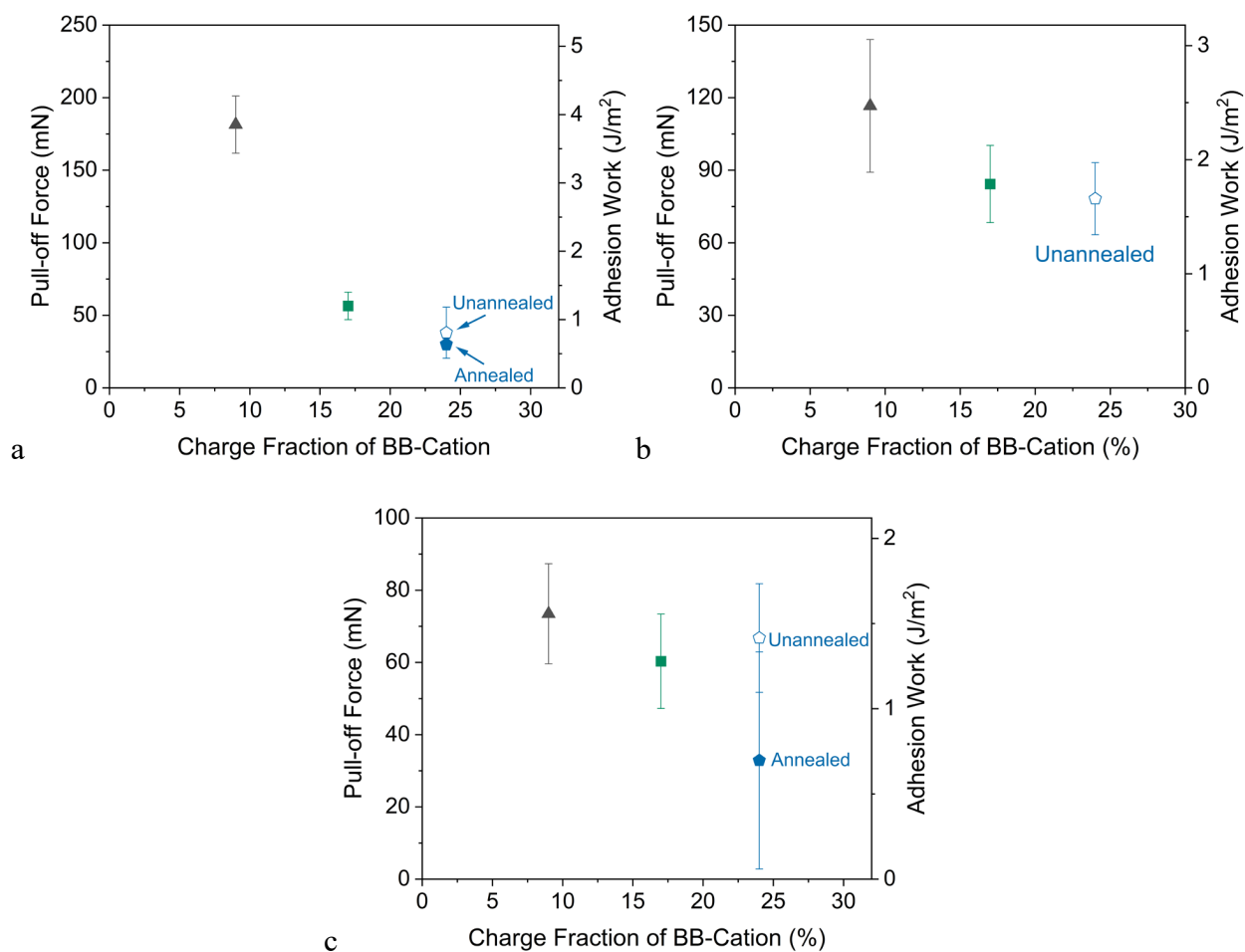

**Fig. S33.** Intrinsic adhesion of BB-Cation with different charge fractions against (a) Linear-Anion, (b) BB-Anion ($\alpha$ = 43%), and (c) BB-Anion ($\alpha$ = 51%).

The normalized adhesion work (on/off ratio) of BB-Cation vs. Linear-Anion, BB-Cation vs. 43% charge fraction BB-Anion and BB-Cation vs. 51% charge fraction BB-Anion was plotted in Fig. S34a, b and c respectively. It should be noted that for BB-Cation at $\alpha$ = 24%, the high viscosity of the uncured precursor made the blade-coated films prone to coating defects and promoted air-bubble entrapment at the surface during blade-coating and subsequent UV crosslinking. These surface imperfections degraded interfacial contact, reducing adhesion and suppressing the on/off ratio ($\alpha$ = 24%, unannealed; Fig. S34a–c). To address this issue, we introduced a post-treatment step in which the BB-Cation film was annealed after blade-coating (prior to crosslinking) at 80 °C under vacuum for 24 h to enable sufficient degassing and improve surface quality (Supplementary Note 10). This treatment yielded markedly improved surface integrity and significantly enhanced normalized adhesion ($\alpha$ = 24%, annealed; Fig. S36a–c). The results were normalized by the intrinsic adhesion in the absence of an applied electric field. Normalized adhesion represents the ratio of adhesive strength when the voltage is turned on vs. off (the "on/off ratio").

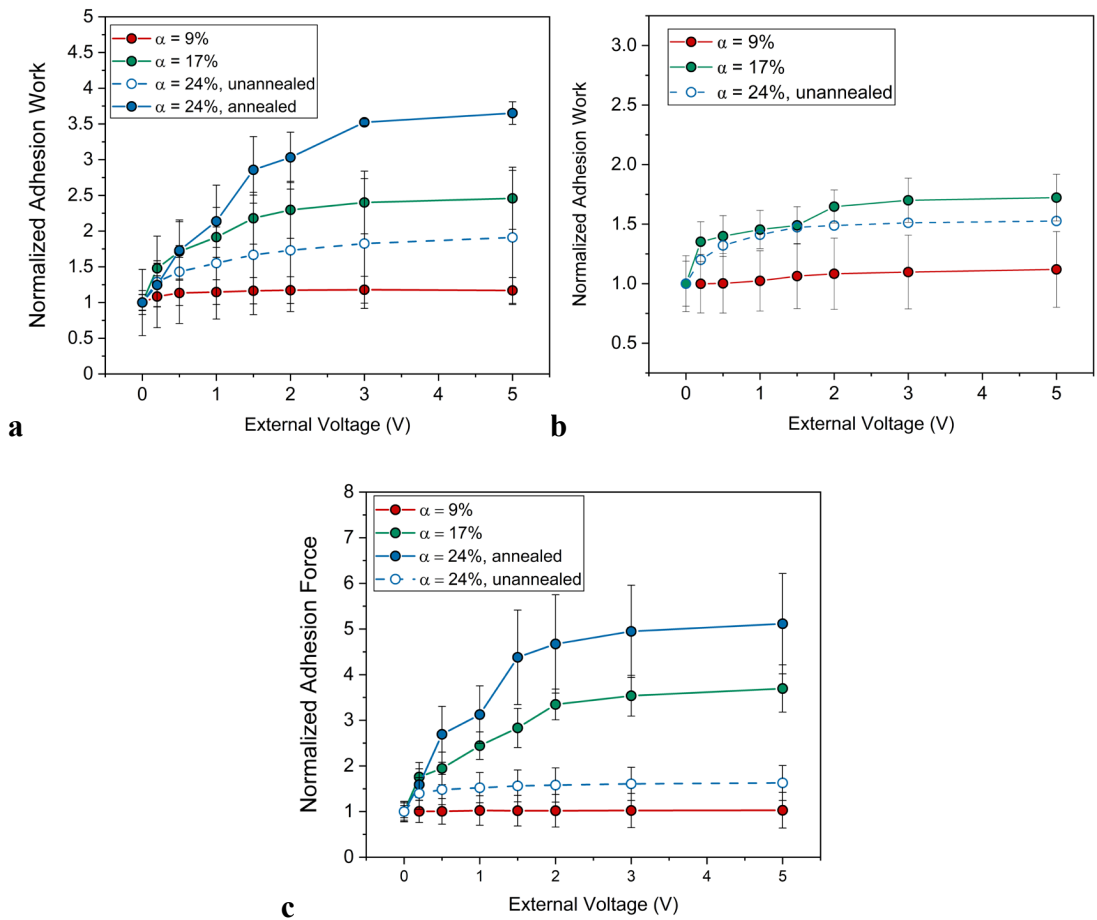

**Fig. S34.** Normalized electro-adhesion of BB-Cation with different charge fractions against (a) Linear-Anion, (b) BB-Anion ($\alpha$ = 43%), and (c) BB-Anion ($\alpha$ = 51%).

## 9. Dwelling time of adhesion test

Figure S35 shows the force versus time for the adhesion test between 17% charge fraction BB-Cation vs. 51% charge fraction BB-Anion with different dwelling times. The adhesion peak force first increases with dwelling time and reaches a plateau value when the dwelling time is more than 300 s. This confirms that 300 s is an appropriate choice of dwelling time for the adhesion tests.

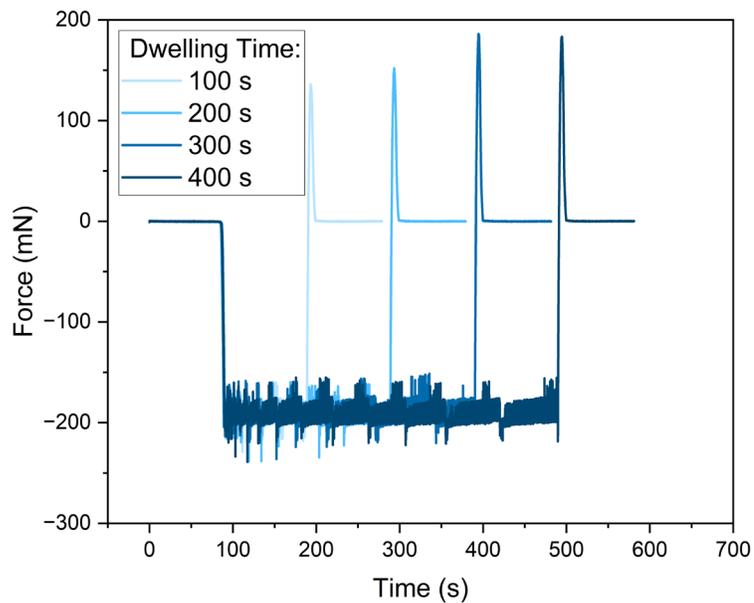

**Fig. S35.** Raw data of adhesion test of 17% charge fraction BB-Cation vs. 51% charge fraction BB-Anion with different dwelling times. We find that further increasing dwelling time above 300 s has no effect on the measured adhesion force.

## 10. Surface Morphology of BB-Cation with different charge fractions

As is mentioned in Supplementary Note 8, BB-Cation at $\alpha = 24\%$ was annealed to improve surface quality. The comparison of BB-Cation at $\alpha = 24\%$ surface before (Fig. S36a and b) and after annealing (Fig. S36d) is shown in Fig. S36. The surface of BB-Cation at $\alpha = 17\%$ is also shown in Fig. S36c for reference.

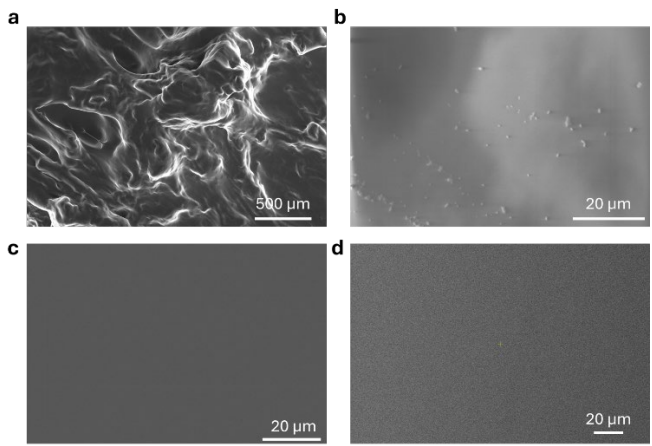

**Fig. S36.** SEM images of unannealed BB-Cation with 24% charge fraction (**a** and **b**), 17% charge fraction (**c**) and annealed BB-Cation with 24% charge fraction (**d**). The unannealed 24% specimen exhibits pronounced surface roughness with entrapped bubbles/voids; the 17% specimen (**c**) and annealed BB-Cation with 30% charge fraction (**d**) are shown for comparison.

## 11. Charge density for BB-Cation and BB-Anion

Tables S5–S7 summarize the charge densities of BB-Cation, BB-Anion, and Linear-Anion, respectively. The actual molecular weight was calculated based on the measured charge fraction. Taking BB-Cation with a nominal charge fraction of 10% ($\alpha = 9\%$) as an example, the nominal molecular weight is calculated as $90 \times$ the molecular weight of the macromonomer plus $10 \times$ the molecular weight of the ionic monomer, whereas the actual molecular weight is calculated as $91 \times$ the molecular weight of the macromonomer plus $9 \times$ the molecular weight of the ionic monomer. Since BB-Cation and BB-Anion precursors are very viscous and thus their volume is difficult to measure, we assume that the density of the elastomer is 1 g/cm$^3$[10], and then the charge

density (C/g) can be converted to charge concentration (C/cm$^3$). For example, the charge concentration of 24% charge fraction BB-Cation is 18.5 C/cm$^3$.

| Nominal Charge Fraction (%) | Actual Charge Fraction (%) | Nominal Molecular Weight (Da) | Actual Molecular Weight (Da) | Charge Amount (C/mol polymer) | Charge Density (C/g) |
|---|---|---|---|---|---|
| 0 | 0 | 150000 | 150000 | 0 | 0 |
| 10 | 9 | 139740 | 140766 | 868369 | 6.2 |
| 20 | 17 | 129480 | 132558 | 1640233 | 12.4 |
| 30 | 24 | 119220 | 125376 | 2315623 | 18.5 |

**Table S5.** Summary of charge density for BB-Cation.

| Nominal Charge Fraction (%) | Actual Charge Fraction (%) | Nominal Molecular Weight (Da) | Actual Molecular Weight (Da) | Charge Amount (C/ mol polymer) | Charge per Mass (C/g) |
|---|---|---|---|---|---|
| 40 | 43 | 322920 | 308139 | 4148824 | 13.5 |
| 50 | 51 | 273650 | 268723 | 4920698 | 18.3 |

**Table S6.** Summary of charge density for BB-Anion.

| Nominal Charge Fraction (%) | Actual Charge Fraction (%) | Molecular Weight of Monomer (Da) | Average Charge Amount of Monomer (C/mol monomer) | Charge per Mass (C/g) |
|---|---|---|---|---|
| 80 | 80 | 223 | 77188 | 346 |

**Table S7.** Summary of charge density for Linear-Anion.